\newcommand\apjcls{1}
\newcommand\aastexcls{2}
\newcommand\othercls{3}

\newcommand\papercls{\aastexcls}
\documentclass[tighten, times, twocolumn, trackchanges]{aastex62}

\if\papercls \apjcls
\usepackage{apjfonts}
\else\if\papercls \othercls
\usepackage{epsfig}
\fi\fi
\usepackage{ifthen}
\usepackage{natbib}
\usepackage{amssymb, amsmath}
\usepackage{appendix}
\usepackage{etoolbox}
\usepackage[T1]{fontenc}
\usepackage{paralist}
\usepackage[normalem]{ulem}
\usepackage{ragged2e}

\if\papercls \apjcls
\newcommand\aas{\ref@jnl{AAS Meeting Abstracts}}
\newcommand\dps{\ref@jnl{AAS/DPS Meeting Abstracts}}

\newcommand\maps{\ref@jnl{MAPS}}
\else\if\papercls \othercls
\usepackage{astjnlabbrev-jh}
\fi\fi

\if\papercls \aastexcls
\hypersetup{citecolor=blue, 
            linkcolor=blue, 
            menucolor=blue, 
            urlcolor=blue}  
\else
\usepackage[
bookmarks=true,           
bookmarksnumbered=true,   
colorlinks=true,          
citecolor=blue,           
linkcolor=blue,           
menucolor=blue,           
urlcolor=blue,            
linkbordercolor={0 0 1},  
pdfborder={0 0 1},
frenchlinks=true]{hyperref}
\fi

\if\papercls \othercls

\else

\fi

\providecommand{\adsurl}[1]{\href{#1}{ADS}}

\makeatletter
\patchcmd{\NAT@citex}
  {\@citea\NAT@hyper@{%
     \NAT@nmfmt{\NAT@nm}%
     \hyper@natlinkbreak{\NAT@aysep\NAT@spacechar}{\@citeb\@extra@b@citeb}%
     \NAT@date}}
  {\@citea\NAT@nmfmt{\NAT@nm}%
   \NAT@aysep\NAT@spacechar\NAT@hyper@{\NAT@date}}{}{}

\patchcmd{\NAT@citex}
  {\@citea\NAT@hyper@{%
     \NAT@nmfmt{\NAT@nm}%
     \hyper@natlinkbreak{\NAT@spacechar\NAT@@open\if*#1*\else#1\NAT@spacechar\fi}%
       {\@citeb\@extra@b@citeb}%
     \NAT@date}}
  {\@citea\NAT@nmfmt{\NAT@nm}%
   \NAT@spacechar\NAT@@open\if*#1*\else#1\NAT@spacechar\fi\NAT@hyper@{\NAT@date}}
  {}{}
\makeatother

\makeatletter
\DeclareRobustCommand{\lowcase}[1]{\@lowcase#1\@nil}
\def\@lowcase#1\@nil{\if\relax#1\relax\else\MakeLowercase{#1}\fi}
\makeatother

\DeclareSymbolFont{UPM}{U}{eur}{m}{n}
\DeclareMathSymbol{\umu}{0}{UPM}{"16}
\let\oldumu=\umu
\renewcommand\umu{\ifmmode\oldumu\else\math{\oldumu}\fi}
\newcommand\micro{\umu}
\if\papercls \othercls
\newcommand\micron{\micro m}
\else
\renewcommand\micron{\micro m}
\fi
\newcommand\microns{\micron}

\let\oldsim=\sim
\renewcommand\sim{\ifmmode\oldsim\else\math{\oldsim}\fi}
\let\oldpm=\pm
\renewcommand\pm{\ifmmode\oldpm\else\math{\oldpm}\fi}
\newcommand\by{\ifmmode\times\else\math{\times}\fi}
\newcommand\ttt[1]{10\sp{#1}}
\newcommand\tttt[1]{\by\ttt{#1}}

\newbox{\wdbox}
\newcommand\cw{\setbox\wdbox=\hbox{,}\hspace{\wd\wdbox}}
\newcommand\iw{\setbox\wdbox=\hbox{i}\hspace{\wd\wdbox}}

\newcount\timect
\newcount\hourct
\newcount\minct
\newcommand\now{\timect=\time \divide\timect by 60
         \hourct=\timect \multiply\hourct by 60
         \minct=\time \advance\minct by -\hourct
         \number\timect:\ifnum \minct < 10 0\fi\number\minct}

\newcommand\mctc{\multicolumn{2}{c}}

\def\hyph{-\penalty0\hskip0pt\relax}

\catcode`@=11
\newcommand\comment[1]{}
\newcommand\commenton{\catcode`\%=14}

\renewcommand\math[1]{$#1$}
\newcommand\mathshifton{\catcode`\$=3}

\let\atab=&
\newcommand\atabon{\catcode`\&=4}

\let\oldmsp=\sp
\let\oldmsb=\sb
\def\sp#1{\ifmmode
           \oldmsp{#1}%
         \else\strut\raise.85ex\hbox{\scriptsize #1}\fi}
\def\sb#1{\ifmmode
           \oldmsb{#1}%
         \else\strut\raise-.54ex\hbox{\scriptsize #1}\fi}
\newbox\@sp
\newbox\@sb
\def\sbp#1#2{\ifmmode%
           \oldmsb{#1}\oldmsp{#2}%
         \else
           \setbox\@sb=\hbox{\sb{#1}}%
           \setbox\@sp=\hbox{\sp{#2}}%
           \rlap{\copy\@sb}\copy\@sp
           \ifdim \wd\@sb >\wd\@sp
             \hskip -\wd\@sp \hskip \wd\@sb
           \fi
        \fi}
\def\msp#1{\ifmmode
           \oldmsp{#1}
         \else \math{\oldmsp{#1}}\fi}
\def\msb#1{\ifmmode
           \oldmsb{#1}
         \else \math{\oldmsb{#1}}\fi}

\def\supon{\catcode`\^=7}

\def\subon{\catcode`\_=8}

\def\supsubon{\supon \subon}

\newcommand\actcharon{\catcode`\~=13}

\newcommand\paramon{\catcode`\#=6}

\comment{And now to turn us totally on and off...}

\newcommand\reservedcharson{ \commenton  \mathshifton  \atabon  \supsubon 
                             \actcharon  \paramon}

\catcode`@=12
\reservedcharson

\if\papercls \apjcls

\else

\fi

\newcommand\SST{{\em Spitzer Space Telescope}}
\newcommand\Spitzer{{\em Spitzer}}
\newcommand\HubbleST{{\em Hubble Space Telescope}}
\newcommand\HST{{\em HST}}

\newcommand\JWST{{\em JWST}}
\newcommand\Webb{{\em James Webb Space Telescope}}
\newcommand\chisq{\ifmmode{\chi\sp{2}}\else\math{\chi\sp{2}}\fi}
\newcommand\redchisq{\ifmmode{ \chi\sp{2}\sb{\rm red}}
                    \else\math{\chi\sp{2}\sb{\rm red}}\fi}
\newcommand\Teq{\ifmmode{T\sb{\rm eq}}\else$T$\sb{eq}\fi}
\newcommand\Tb{\ifmmode{T\sb{\rm b}}\else$T$\sb{b}\fi}
\newcommand\mjup{\ifmmode{M\sb{\rm Jup}}\else$M$\sb{Jup}\fi}
\newcommand\rjup{\ifmmode{R\sb{\rm Jup}}\else$R$\sb{Jup}\fi}
\newcommand\msun{\ifmmode{M\sb{\odot}}\else$M\sb{\odot}$\fi}
\newcommand\rsun{\ifmmode{R\sb{\odot}}\else$R\sb{\odot}$\fi}
\newcommand\Rs{\ifmmode{R\sb{\rm s}}\else$R\sb{\rm s}$\fi}
\newcommand\mearth{\ifmmode{M\sb{\oplus}}\else$M\sb{\oplus}$\fi}
\newcommand\rearth{\ifmmode{R\sb{\oplus}}\else$R\sb{\oplus}$\fi}
\newcommand\Rp{\ifmmode{R\sb{\rm p}}\else$R\sb{\rm p}$\fi}

\newcommand\molhyd{\ifmmode{{\rm H}\sb{2}}\else{H$\sb{2}$}\fi}
\newcommand\methane{\ifmmode{{\rm CH}\sb{4}}\else{CH$\sb{4}$}\fi}
\newcommand\water{\ifmmode{{\rm H}\sb{2}{\rm O}}\else{H$\sb{2}$O}\fi}
\newcommand\carbdiox{\ifmmode{{\rm CO}\sb{2}}\else{CO$\sb{2}$}\fi}
\newcommand\carbmono{\ifmmode{{\rm CO}}\else{CO}\fi}
\newcommand\ammonia{\ifmmode{{\rm NH}\sb{3}}\else{NH$\sb{3}$}\fi}
\newcommand\acetylene{\ifmmode{{\rm C}\sb{2}{\rm H}\sb{2}}
                        \else{C$\sb{2}$H$\sb{2}$}\fi}
\newcommand\ethylene{\ifmmode{{\rm C}\sb{2}{\rm H}\sb{4}}
                        \else{C$\sb{2}$H$\sb{4}$}\fi}
\newcommand\cyanide{\ifmmode{{\rm HCN}}\else{HCN}\fi}
\newcommand\nitrogen{\ifmmode{{\rm N}\sb{2}}\else{N$\sb{2}$}\fi}

\newcommand\repack{\textsc{repack}}
\newcommand\pyratbay{\textsc{Pyrat Bay}}

\newcommand\mcc{\textsc{mc3}}
\newcommand\pandexo{\textsc{PandExo}}

\newcommand\pcloud{\ifmmode{p\sb{\rm cloud}}
                 \else\math{p\sb{\rm cloud}}\fi}

\newcommand\der{\ifmmode{\rm d}\else\math{\rm d}\fi}

\NoNewPageAfterKeywords

\newcommand\disk{disk{\hyph}integrated}
\newcommand\resolved{longitudinally resolved}
\newcommand\xwater{\ifmmode{X_{\rm H2O}}\else{$X_{\rm H2O}$}\fi}
\newcommand\xmethane{\ifmmode{X_{\rm CH4}}\else{$X_{\rm CH4}$}\fi}
\newcommand\xcarbmono{\ifmmode{X_{\rm CO}}\else{$X_{\rm CO}$}\fi}
\newcommand\xcarbdiox{\ifmmode{X_{\rm CO2}}\else{$X_{\rm CO2}$}\fi}

\shorttitle{Longitudinally Resolved Spectral Retrieval}
\shortauthors{Cubillos {\em et al.}}

\begin{document}
\title{Longitudinally Resolved Spectral Retrieval (ReSpect) of WASP-43b}

\author[0000-0002-1347-2600]{Patricio E. Cubillos}
\affiliation{Space Research Institute, Austrian Academy of Sciences,
             Schmiedlstrasse 6, A-8042, Graz, Austria}

\author[0000-0001-9887-4117]{Dylan Keating}
\affiliation{Department of Physics, McGill University, 3600 rue University, Montr\'eal, QC H3A 2T8, Canada}

\author[0000-0001-6129-5699]{Nicolas B.~Cowan}
\affiliation{Department of Physics, McGill University, 3600 rue University, Montr\'eal, QC H3A 2T8, Canada}
\affiliation{Department of Earth \& Planetary Sciences, McGill University, 3450 rue University, Montr\'eal, QC H3A 0E8, Canada}

\author[0000-0003-0489-1528]{Johanna M. Vos}
\affiliation{Department of Astrophysics, American Museum of Natural History, 200 Central Park West, New York, NY 10024, USA}

\author[0000-0003-4600-5627]{Ben Burningham}
\affiliation{Centre for Astrophysics Research, Department of Physics, Astronomy and Mathematics, University of Hertfordshire, Hatfield AL10 9AB, UK}

\author[0000-0001-7591-2731]{Marie Ygouf}
\affiliation{Jet Propulsion Laboratory, California Institute of Technology, Pasadena, CA 91109, USA}

\author[0000-0001-7356-6652]{Theodora Karalidi}
\affiliation{Department of Physics, University of Central Florida, 4111 Libra Dr, Orlando, FL, 32816, USA}

\author{Yifan Zhou}
\affiliation{Department of Astronomy, The University of Texas at Austin, Austin, TX 78712, USA}
\affiliation{McDonald Observatory, The University of Texas, Austin, TX 78712, USA}

\author[0000-0003-4636-6676]{Eileen C. Gonzales}
\altaffiliation{51 Pegasi b Fellow}
\altaffiliation{LSSTC Data Science Fellow}
\affiliation{Department of Astronomy and Carl Sagan Institute, Cornell University, 122 Sciences Drive, Ithaca, NY 14853, USA}
\affiliation{Department of Astrophysics, American Museum of Natural History, 200 Central Park West, New York, NY 10024, USA}
\affiliation{The Graduate Center, City University of New York, New York, NY 10016, USA}
\affiliation{Department of Physics and Astronomy, Hunter College, City University of New York, New York, NY 10065, USA}

\email{patricio.cubillos@oeaw.ac.at}

\begin{abstract}
Thermal phase variations of short period planets indicate that they
are not spherical cows: day-to-night temperature contrasts range from
hundreds to thousands of degrees, rivaling their vertical temperature
contrasts.  Nonetheless, the emergent spectra of short-period planets
have typically been fit using one-dimensional (1D) spectral retrieval
codes that only account for vertical temperature gradients. The
popularity of 1D spectral retrieval codes is easy to understand: they
are robust and have a rich legacy in Solar System atmospheric
studies. Exoplanet researchers have recently introduced
multi-dimensional retrieval schemes for interpreting the spectra of
short-period planets, but these codes are necessarily more complex and
computationally expensive than their 1D counterparts.  In this paper
we present an alternative: phase-dependent spectral observations are
inverted to produce longitudinally resolved spectra that can then be
fitted using standard 1D spectral retrieval codes.  We test this
scheme on the iconic phase-resolved spectra of WASP-43b and on
simulated {\JWST} observations using the open-source {\pyratbay} 1D
spectral retrieval framework.  Notably, we take the model complexity
of the simulations one step further over previous studies by allowing
for longitudinal variations in composition in addition to temperature.
We show that performing 1D spectral retrieval on longitudinally
resolved spectra is more accurate than applying 1D spectral retrieval
codes to {\disk} emission spectra, despite being identical in terms of
computational load. We find that for the extant \emph{Hubble}
and \emph{Spitzer} observations of WASP-43b the difference between the
two approaches is negligible but that {\JWST} phase measurements
should be treated with
longitudinally \textbf{re}solved \textbf{spect}ral retrieval
(ReSpect).
\end{abstract}

\section{Introduction}

Unlike Solar System worlds, short-period planets are poorly modeled by
one-dimensional (1D) atmospheric models.  Tidal forces tend to lock
short-period planets into synchronous rotation with their host star,
while the greater incident flux and resulting high temperatures lead
to short radiative
timescales \citep{ShowmanGuillot2002aapCirctide51Pegb}.  As a result,
the atmospheric composition---and even phase---can differ
qualitatively between a planet's dayside and nightside, e.g., an ultra
hot Jupiter will have atomic gas on its dayside and molecular gas on
its nightside \citep{BellCowan2018apjUltraHotHeatTransport,
TanKomacek2019apjUHJcirculation}, while a lava planet has vaporized
rock on its dayside but an airless
nightside \citep{LegerEtal2011icarCorot7b}.  Even ``modest''
temperature contrasts of a few hundred degrees between day and night
might lead to first order shifts in chemistry \citep[CH$_4$ vs.\
CO;][]{CooperShowman2008apjDynamicsAndChemistry,
AgundezEtal2012aaChemistry} and aerosols \citep[clouds on the
nightside, but clear skies on the
dayside;][]{ParmentierEtal2016apjInhomogeneousClouds,
ParmentierEtal2021mnrasCloudyPhaseCurves,
RomanEtal2021apj3DcloudyPhaseCurves}.  For relatively cool and rapidly
spinning worlds in the Solar System, a 1D ``spherical cow''
atmospheric model adequately captures the dominant vertical
temperature gradients.  Short period planets instead need to be
considered as at least two-dimensional objects (altitude and
longitude).  We cannot claim to really understand a short period
planet's atmosphere unless we have explained its day-to-night
differences.

The most direct way to observe a planet's day-to-night variations in
temperature, composition, and aerosols is via phase-curve
measurements. Photometric thermal phase variations have been available
for more than a decade \citep{KnutsonEtal2007nat189733Phase} and
enable us to build longitudinal brightness
maps \citep{CowanAgol2008apjPhaseInversion} for dozens of hot Jupiters
and a few smaller planets \citep[for a recent review
see][]{ParmentierCrossfield2018haexPhaseCurves}. Longitudinal
``temperature maps'' inferred from photometric phase variations should
be interpreted with a grain of salt, however: given the significant
changes in composition and aerosols, the opacities can vary
dramatically as a function of longitude, so the brightness map in a
single band may not represent bolometric flux, nor the temperature
along an
isobar \citep[][]{DobbsDixonCowan2017apjHotJupiterMapping}. The only
way to truly understand the longitudinal variations of a short-period
planet are multi-band phase-curve
measurements \citep{KnutsonEtal2009apjHD189phases,
KnutsonEtal2012apjHD189phaseCH1CH2}.

{\cite{StevensonEtal2014sciWASP43bHSTphase,
StevensonEtal2017ajWASP43bSpitzerPhase}} presented 1D spectral
retrievals based on {\disk} spectroscopy of the hot Jupiter WASP-43b
at four orbital phases. This approach is not entirely self-consistent,
however: 1D spectral retrieval is predicated on a horizontally uniform
planet, whereas the time-varying {\disk} brightness is a testament to
the planet's longitudinal
inhomogeneity \citep[e.g.,][]{FengEtal2016apjNonUniformTempProfiles,
IrwinEtal2020mnras2.5Dretrieval,
CaldasEtal2019aaTransmission3Deffects,
TaylorEtal2020mnrasEmissionBiases}.  Multidimensional retrievals of
the WASP-43b phase-curve data
by \citet{IrwinEtal2020mnras2.5Dretrieval}
and \citet{FengEtal2020aj2Dretrievals} have shown significant
deviations from previous 1D retrieval results.

\subsection{Thermal-emission Spectral Retrieval for Inhomogeneous Planets}

Recently, there has been a growing interest in performing spectral
retrievals for planets with inhomogeneous atmospheres, prompted by the
higher quality data that next-generation observatories will
provide.

{\citet{FengEtal2016apjNonUniformTempProfiles}} presented the first
exoplanet atmospheric retrievals considering models beyond the 1D
assumption.  They modeled the planetary flux as a linear combination
arising from a ``hot'' and a ``cold'' thermal profile component
(the components of this 2TP model are averaged and thus represent
an observation at quadrature).  They
showed that the assumption of a single 1D thermal profile biases the
retrieved composition of a hot Jupiter atmosphere that is composed of
two thermal profiles.  When there is a strong day--night temperature
contrast, the 1D model overestimated the {\methane} abundance, 
whereas the two-component model yielded
an upper limit, consistent with the input {\methane} abundance.  They
found a similar result when applying the analysis to the WASP-43b
phase-curve data at orbital phase = 0.25.

{\cite{IrwinEtal2020mnras2.5Dretrieval}} presented ``2.5-dimensional''
spectral retrievals of WASP-43b using optimal
estimation \citep[e.g.,][]{Rodgers2000bookInverseMethodsAtmosphericSounding}.
They simultaneously fit the {\disk} spectra at many different orbital
phases with an atmospheric model that is a discrete function of
longitude and an assumed latitudinal dependence. The number of
parameters is therefore proportional to the chosen number of
longitudinal slices, $N_l=16$.  Since multiple longitudinal slices
contribute to the {\disk} spectrum at a given phase, each slice is
constrained by data from multiple phases. This approach has the
advantage of accounting for the different viewing angles for regions
near the center of the planetary disk vs.\ near the limb of the
planet---including the poles. Moreover, the atmospheric temperature
and composition can vary arbitrarily from one longitudinal slice to
the next, which makes this approach very flexible.

{\cite{TaylorEtal2020mnrasEmissionBiases}} quantified the impact of
performing a 1D spectral retrieval on {\disk} observations of a
horizontally inhomogeneous planet. They note that short-wavelength
data are particularly useful at diagnosing a mixture of planetary
regions with different temperatures---this is likely a consequence of
the stronger temperature-dependence on the Wien than on the
Rayleigh-Jeans side of the Planck function \citep[cf.\ the
sum-of-blackbodies
of][]{SchwartzCowan2015mnrasAlbedoEnergyBudget}. The authors first
develop an admixture of two 1D
models \citep[following][]{FengEtal2016apjNonUniformTempProfiles}
before showing that diluting a single hot temperature--pressure
profile with a region that emits no flux whatsoever adequately fits
synthetic emission data.

{\citet{FengEtal2020aj2Dretrievals}} extended their 2TP
approach \citep{FengEtal2016apjNonUniformTempProfiles} to model a
planet's phase curve-emission at any given orbital phase by weighting
the intensity from the hot and cold components according to
appropriate viewing-geometry corrections.  Their analysis of synthetic
phase curves of WASP-43b with globally constant composition but
inhomogeneous day and night temperature profiles showed that 1D
retrievals of simulated 2TP-type planets can significantly
overestimate the abundances of species that are absent from
the atmosphere (e.g., {\methane}).  Their analysis of observed
WASP-43b phase curves with the 2TP approach tends to favor upper
limits for the {\methane} abundance at a few more orbital phases than
the 1D approach.

\subsection{Our Approach: Spectral Mapping}

Our approach is complementary to the efforts described above: rather
than make the forward model more complex by adding dimensions or
atmospheric columns, we opt to process the data one step further than
previous researchers. We convert the time-resolved spectra into
longitudinally resolved spectra by applying the analytic formalism
of \cite{CowanAgol2008apjPhaseInversion} at each wavelength.  These
spectra can then be interpreted with 1D spectral retrieval codes.  

It has long been recognized that time-resolved multi-band photometry
can be converted to longitudinal
maps \citep[e.g.,][]{KnutsonEtal2009apjHD189phases,
KnutsonEtal2012apjHD189phaseCH1CH2}, we simply extend this approach to
spectral data. In principle, our approach should yield more accurate
atmospheric retrievals than the traditional approach of performing 1D
spectral retrieval on {\disk} spectra, while side-stepping the
development of higher-dimensional retrieval codes.

In \S\ref{sec:retrieval} we apply our approach to the iconic spectral
phase curve measurements of
WASP-43b \citep{StevensonEtal2014sciWASP43bHSTphase,
StevensonEtal2017ajWASP43bSpitzerPhase} and in \S\ref{sec:synthetic}
to simulated {\JWST} observations of the same planet.  We discuss our
results in \S\ref{sec:discussion} and conclude
in \S\ref{sec:conclusions}.

\section{Atmospheric Retrieval}
\label{sec:retrieval}

In this section we retrieve existing spectral phase-curve observations
of the hot-Jupiter planet WASP-43b.  This study serves multiple
purposes.  First, it allows us to compare the standard 1D {\disk}
retrieval approach to previous studies from the literature.  Second,
it allow us to compare the {\resolved} retrieval to the {\disk}
approach under current observational capabilities.

\subsection{WASP-43b}

WASP-43b \citep{HellierEtal2011aaWASP43bdisc} is a
highly{\hyph}irradiated hot{\hyph}Jupiter planet ($R_{\rm p} =
1.04$~\rjup, $M_{\rm p}=2.03$ \mjup) orbiting a K7 dwarf star ($R_{\rm
s}=0.67$~$R_\odot$, $T_{\rm eff}=4520$~K) in a short 19.5 hr
orbit \citep{GillonEtal2012aaTrappistWASP43b}.  These favorable system
properties prompted several atmospheric characterization efforts via
secondary-eclipse and phase-curve observations.  Infrared emission
observations during eclipse with the {\SST} ruled out the presence of
a strong thermal inversion \citep{BlecicEtal2014apjWASP43b}, whereas
optical observations with the {\HubbleST} ({\HST}) detected the water
1.4~{\micron} feature, consistent with solar-composition
values \citep{KreidbergEtal2014apjWASP43bWFC3}.  Full{\hyph}orbit
phase{\hyph}curve observations with HST and
Spitzer \citep{StevensonEtal2014sciWASP43bHSTphase,
StevensonEtal2017ajWASP43bSpitzerPhase} suggest a high day--night
temperature contrast of $\sim$1000~K, hence weak heat redistribution,
but accompanied by significant emission asymmetry, with phase curves
peaking $\sim$40 minutes before the secondary eclipse.  3D atmospheric
circulation models of the planet that exhibit an equatorial
superrotating jet that predict the observed eastward-shifted
hotspot \citep{KatariaEtal2015apjWASP43bGCM} but not the very low
nightside flux that has been contested in subsequent reanalyses of the
data \citep{KeatingCowan2017apjWASP43bHeatTransport,
LoudenEtal2018mnrasSpiderman, MendoncaEtal2018ajWASP43bPhase}.

{\renewcommand{\arraystretch}{1.05}
\begin{table*}[tb]
\centering
\caption{{\pyratbay} Atmospheric Retrievals of WASP43b}
\label{table:WASP43b_retrieval}
\begin{tabular*}{\linewidth} {@{\extracolsep{\fill}} lccccccc}
\hline
\hline
                        &                         &  \mctc{Orbital phase 0.25}             & \mctc{Orbital phase 0.5}                & \mctc{Orbital phase 0.75} \\
Parameter               & Prior                   &  Disk              & Resolved          &  Disk              & Resolved           & Disk                & Resolved  \\
\hline
$\log_{10}(\kappa')$    & $\mathcal U(-7.0, 3.0)$  &  $-5.4_{-0.8}^{+1.5}$ & $-5.5_{-0.7}^{+1.4}$ & $-3.4_{-1.4}^{+1.1}$ & $-2.8_{-1.2}^{+1.1}$ & $-4.2_{-1.3}^{+1.6}$ & $-3.8_{-1.4}^{+1.6}$ \\
$\log_{10}(\gamma)$     & $\mathcal U(-4, 4)$      &  $-1.1_{-0.9}^{+0.4}$ & $-1.0_{-0.7}^{+0.4}$ & $-0.9_{-0.8}^{+0.4}$ & $-1.1_{-1.0}^{+0.5}$ & $-1.3_{-1.1}^{+0.6}$ & $-1.4_{-1.1}^{+0.6}$ \\
$T_{\rm irr}$ (K)        & $\mathcal U(100, 3000)$  &  $1310_{-560}^{+155}$ & $1360_{-460}^{+125}$ & $1300_{-490}^{+275}$ & $1175_{-510}^{+360}$ & $1000_{-495}^{+260}$ & $930_{-455}^{+280}$ \\
$\log_{10}(X_{\rm H2O})$  & $\mathcal U(-12, -1)$    & $-4.4_{-0.4}^{+1.6}$ & $-4.2_{-0.4}^{+1.7}$ & $-2.5_{-1.0}^{+0.9}$ & $-2.1_{-0.8}^{+0.7}$ & $-3.8_{-1.0}^{+1.7}$ & $-3.3_{-1.2}^{+1.4}$ \\
$\log_{10}(X_{\rm CO})$  & $\mathcal U(-12, -1)$    & $-7.4_{-3.1}^{+3.9}$ & $-7.4_{-3.2}^{+4.0}$ & $-4.6_{-4.9}^{+2.7}$ & $-4.7_{-4.8}^{+2.8}$ & $-5.4_{-4.4}^{+2.9}$ & $-5.0_{-4.8}^{+2.8}$ \\
$\log_{10}(X_{\rm CO2})$  & $\mathcal U(-12, -1)$    & $-8.0_{-2.8}^{+3.9}$ & $-7.9_{-2.8}^{+4.0}$ & $-5.2_{-3.6}^{+1.4}$ & $-5.3_{-3.5}^{+1.2}$ & $-4.7_{-3.5}^{+2.3}$ & $-4.2_{-3.8}^{+2.0}$ \\
$\log_{10}(X_{\rm CH4})$  & $\mathcal U(-12, -1)$    & $-6.7_{-3.6}^{+2.5}$ & $-6.1_{-3.7}^{+2.1}$ & $-8.4_{-2.5}^{+2.6}$ & $-8.7_{-2.2}^{+2.4}$ & $-6.7_{-3.6}^{+2.4}$ & $-6.8_{-3.6}^{+2.4}$ \\
$\log_{10}(p_{\rm cloud}/{\rm bar})$ & $\mathcal U(-6, 2)$ & $0.9_{-0.6}^{+0.8}$ & $0.8_{-0.7}^{+0.8}$ & $0.1_{-1.3}^{+1.3}$ & $-0.1_{-1.4}^{+1.4}$ & $0.5_{-1.2}^{+1.0}$ & $0.4_{-1.3}^{+1.1}$ \\
\hline
\end{tabular*}
\begin{minipage}{\linewidth}
\vspace{0.1cm}
\footnotesize {\bf Notes.} The reported retrieved values correspond to the
marginal posterior distribution's median and boundaries of the 68\%
central credible interval \citep{Andrae2010arxivErrorEstimation}. \\
\end{minipage}
\end{table*}
}

\subsection{Converting Spectral Phase Curves to Longitudinally Resolved Spectra}
\label{sec:converting}

The thermal phase variations of an exoplanet are, to good
approximation, periodic at the orbital period and hence may be
approximated by a Fourier series.  Sinusoids turn out to be not only a
convenient parameterization for the observed signal, they also lend
themselves to analytic conversion between the phase variations and a
longitudinal brightness map of the
planet \citep{CowanAgol2008apjPhaseInversion,CowanEtal2013mnrasLightCurves}.
Phase-curve mapping suffers from two intrinsic degeneracies: 1) there
is nearly no latitudinal information in the phase curves, and 2)
certain brightness patterns have no lightcurve signature, a
so-called \emph{nullspace} of the transformation \citep[for a review
of exoplanet mapping see][]{CowanFujii2018bookMappingExoplanets}.

For the WASP-43b {\disk} dataset, we selected the {\HST} WFC3
phase-resolved spectra presented in Table 5
of \citet{StevensonEtal2017ajWASP43bSpitzerPhase}, combined with the
broadband {\Spitzer} 3.6~{\micron} and 4.5~{\micron} phase curves
reanalyzed by \citet{MendoncaEtal2018ajWASP43bPhase}.  We inverted the
emission spectra as a function of orbital phase into spectra as a
function of planetary longitude by applying the analytic mapping
formalism of \citet{CowanAgol2008apjPhaseInversion} on the published
phase curve parameters.  We describe how to analytically estimate maps
and their uncertainties for a variety of phase-curve parameterizations
in Appendices \ref{sec:analytics}, \ref{sec:map_diffs}
and \ref{sec:reparameterization}.  Without enforcing positive
brightness maps, we used a 10,000 iteration Monte Carlo to propagate
uncertainties on the published phase curve parameters to the map
parameters, and hence the longitudinally resolved spectra themselves.

\subsection{WASP-43b HST+Spitzer Retrieval Setup}
\label{sec:wasp43_retrieval}

To study the atmosperic properties of WASP-43b we used the open-source
{\pyratbay} framework\footnote{\href{https://pyratbay.readthedocs.io/}
{https://pyratbay.readthedocs.io/}}
for exoplanet atmospheric
modeling, spectral synthesis, and Bayesian
retrieval \citep[][]{CubillosBlecic2021mnrasPyratBay}.  The
{\pyratbay} package provides forward- and retrieval-modeling
capability, enabling the user to adopt a wide range of custom,
self-consistent, or parametric models of the atmospheric temperature,
abundance, and altitude profiles.  The code can compute emission or
transmission spectra considering opacities from molecular line
transitions, collision-induced absorption, Rayleigh scattering, gray
clouds, and alkali resonance lines.  The atmospheric retrieval
explores the parameter space via a differential-evolution Markov-chain
Monte Carlo sampler \citep{terBraak2008SnookerDEMC}, implemented
through the open-source code {\mcc} \citep{CubillosEtal2017apjRednoise},
and monitors the \citet{GelmanRubin1992stascGRstatistics} statistics
for convergence.

We modeled the atmosphere as a set of 61 pressure layers ranging from
100 to $\ttt{-8}$~bar.  We adopted a parametric temperature profile
using the Eddington approximation
model \citep{Guillot2010aaRadiativeEquilibrium,
LineEtal2013apjRetrievalI}.  We kept fixed the channel partitioning at
$\alpha=0$ and the internal temperature at $T_{\rm int} = 0$~K, which
reduces the temperature profile model to $T(p) = T_{\rm
irr} \sqrt[4]{\xi(\tau)/4}$, where
\begin{equation}
\xi(\tau) = 2 + \frac{2}{\gamma}
  \left[1 + \left(\frac{\gamma\tau}{2}-1\right)e^{-\gamma\tau}\right] +
  2\gamma \left(1-\frac{\tau\sp{2}}{2}\right)E\sb{2}(\gamma\tau),
\end{equation}
with $\tau=\kappa'p$, $p$ the atmospheric pressure, and $E_2(x)$ the
second-order exponential integral.  The retrieval parameters of this
model are thus the irradiation temperature of the planet $T_{\rm
irr}$, the visible--thermal ratio of the Planck mean opacities
$\gamma$, and the thermal Planck mean opacity $\kappa'$.  For the
atmospheric composition we retrieve the volume mixing ratios ($X_{i}$)
of {\water}, CO, {\carbdiox}, and {\methane}, assumed to be constant
with altitude.  The remaining composition is assumed to be {\molhyd},
He, and H in solar-abundance ratios under thermochemical
equilibrium \citep{AsplundEtal2009araSolarComposition}.  The altitude
profile is computed using the hydrostatic-equilibrium equation,
calculated consistently according to the composition (i.e., depending
on the mean molecular mass) and temperature of each retrieval sample.

The {\pyratbay} radiative transfer computed the planetary emission
spectrum between 1.0 and 5.5 {\microns}, at a constant resolving power
of $R=\lambda/\Delta\lambda=10,000$.  To compute the WASP-43
planet-to-star flux ratios we adopted the system parameters
from \citet{GillonEtal2012aaTrappistWASP43b} and the Kurucz stellar
emission model \citep{CastelliKurucz2003aiausATLAS9} assuming solar
metallicity, effective temperature $T_{\rm eff}=4500$~K, surface
gravity $\log(g)=4.5$, and planet-to-star radius ratio of $R_{\rm
p}/R_{\rm s}=0.1596$.

The opacities included the HITEMP line-by-line data for
CO \citep{LiEtal2015apjsCOlineList} and
{\carbdiox} \citep{RothmanEtal2010jqsrtHITEMP}, and the ExoMol
line-by-line data for
{\water} \citep{PolyanskyEtal2018mnrasPOKAZATELexomolH2O} and
{\methane} \citep{YurchenkoTennyson2014mnrasExomolCH4}.  Since the
ExoMol line lists consist of several billions of line transitions, we
employed the {\repack} algorithm \citep{Cubillos2017apjRepack} to
extract only the main transitions that dominate the absorption
spectrum, reducing the number of transitions by a factor of
$\sim$100 \citep[][submitted]{Cubillos2021apjRepackExomol}.  The model
also included collision-induced opacities for {\molhyd}--{\molhyd}
pairs \citep{BorysowEtal2001jqsrtH2H2highT, Borysow2002jqsrtH2H2lowT}
and {\molhyd}--He pairs \citep{BorysowEtal1988apjH2HeRT,
BorysowEtal1989apjH2HeRVRT, BorysowFrommhold1989apjH2HeOvertones};
Rayleigh-scattering opacity for
{\molhyd} \citep{LecavelierEtal2008aaRayleighHD189733b}; and a gray
cloud deck parameterized by the cloud top pressure (\pcloud).  Below
this pressure level the atmosphere becomes instantly opaque at all
wavelengths.  Prior to the MCMC run, we sampled the line-by-line
opacities into the wavelength and pressure grids of the atmospheric
model, as wells as over a temperature array evenly spaced from 100~K
to 3000~K with a step of 100~K.  Thus, during the MCMC the code only
interpolates (linearly) in temperature according to the temperature
profile of each iteration.

\begin{figure*}[t]
\centering
\includegraphics[width=\linewidth, clip]{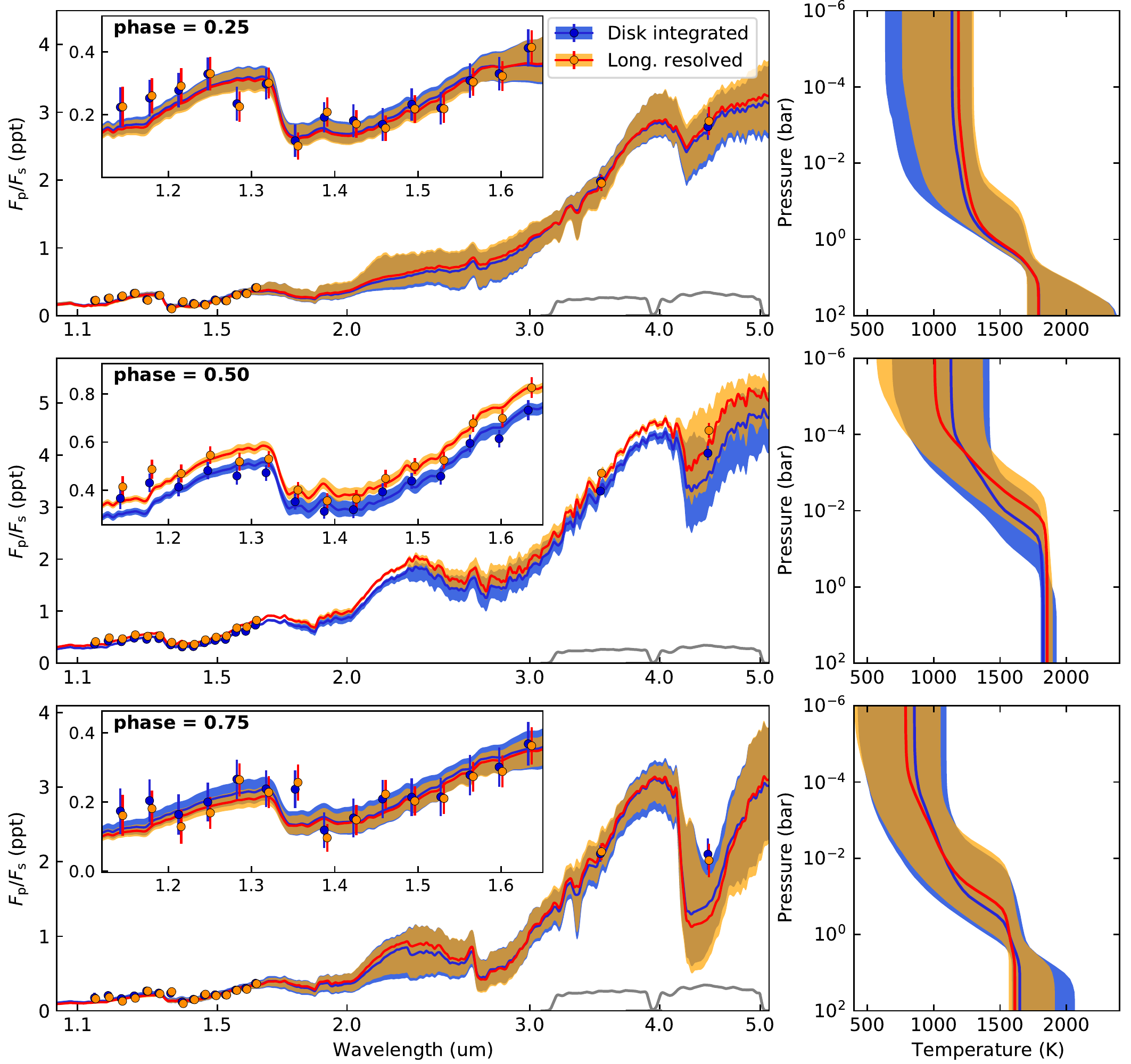}
\caption{
  WASP-43b retrieval with {\pyratbay} of
  the \citet{StevensonEtal2017ajWASP43bSpitzerPhase}
  and \citet{MendoncaEtal2018ajWASP43bPhase} spectra at orbital phases
  of 0.25 (top), 0.5 (middle), and 0.75 (bottom).  All panels follow
  the same color coding, where blue and orange colors
  correspond to the {\disk} and {\resolved} analyses, respectively.
  The solid curves and shaded areas denote the median and 68\%
  credible interval of the posterior distributions for the spectra
  (left panels) and temperature profiles (right panels).  The colored
  markers with error bars show the datasets being fit (left panels).
  The {\resolved} data points have been slightly shifted in wavelength
  for better visibility.  The inset panels zoom in on the wavelengths
  observed by {\HST}.}
\label{fig:WASP43b_stevenson_spectra}
\end{figure*}

Table \ref{table:WASP43b_retrieval} summarizes the retrieval
parameterization, priors, and results of our analysis of
the \citet{StevensonEtal2017ajWASP43bSpitzerPhase}
and \citet{MendoncaEtal2018ajWASP43bPhase} spectra.  In addition to
these priors, the retrieval also required any temperature value to lie
in the range allowed by the opacity data (i.e., $100\ {\rm K} < T(p) <
3000\ {\rm K}$), and the atmosphere to have a primary composition
(i.e., $\sum X_i < 0.1$, for $i \in$ \{{\water}, CO, {\carbdiox},
{\methane}\}).

\begin{figure*}[t]
\centering
\includegraphics[width=\linewidth, clip]{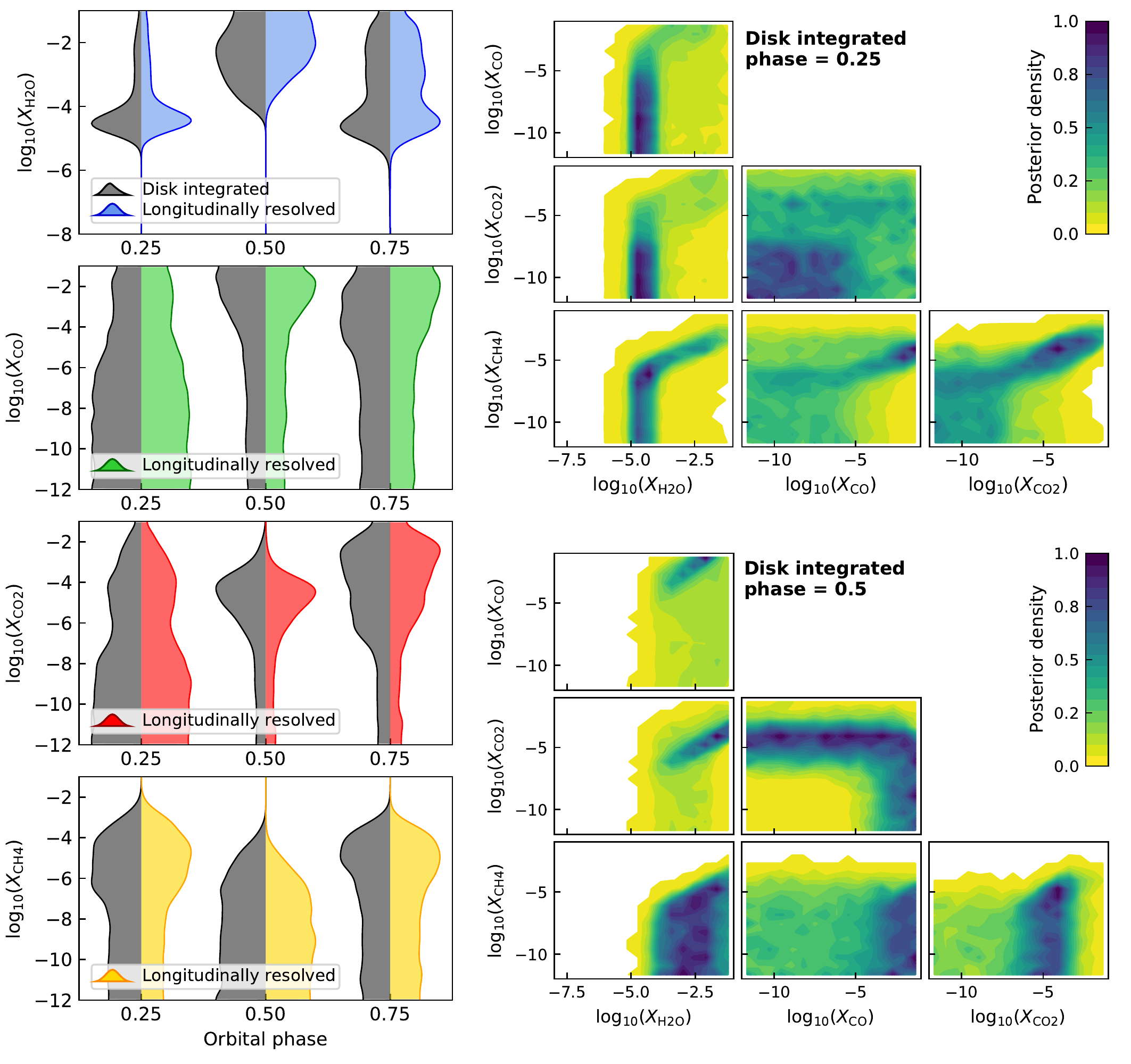}
\caption{
  Retrieved WASP-43b volume-mixing ratios of
  the \citet{StevensonEtal2017ajWASP43bSpitzerPhase}
  and \citet{MendoncaEtal2018ajWASP43bPhase} phase-curve spectra.  The
  left panels show the abundance marginal posterior distributions for
  the {\resolved} (colored) and {\disk} (gray) analyses at the three
  orbital phases shown in Fig.~\ref{fig:WASP43b_stevenson_spectra}.
  The marginal posteriors have been smoothed for better visualization.
  The sets of panels on the right show two examples of the abundance
  pair-wise posterior distributions at orbital phases of 0.25 (top
  panels) and 0.5 (bottom panels) of the {\disk} analysis.}
\label{fig:WASP43b_stevenson_composition}
\end{figure*}

\begin{figure*}[t]
\centering
\includegraphics[width=\linewidth, clip]{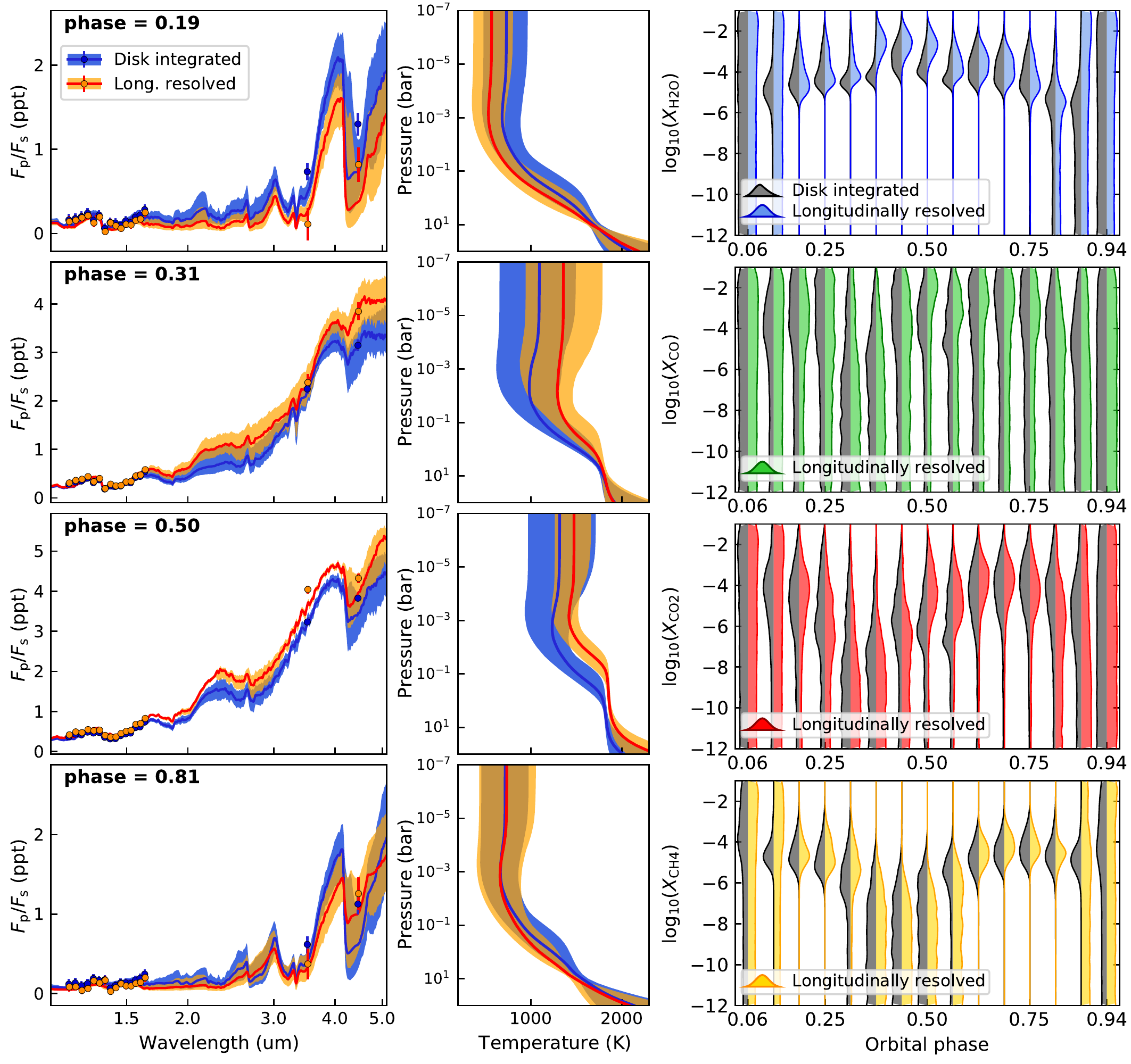}
\caption{
  WASP-43b retrieval with {\pyratbay} of the
  \citet{StevensonEtal2017ajWASP43bSpitzerPhase} phase-curve spectra,
  following the atmospheric parameterization of
  \citet{FengEtal2020aj2Dretrievals}.  The left and center panels
  shows the retrieved spectra and temperature profiles at selected
  phases (see labels) from top to bottom.  The solid curves and shaded areas
  denote the median and 68\% credible interval of the posterior
  distributions for the {\disk} (blue colors) and {\resolved} (orange
  colors) analyses.  The colored markers with error bars show the
  datasets being fit (left panels).
  The right panels show the abundance marginal posterior distributions
  for the {\disk} (gray) and {\resolved} (colored) analyses as a
  function of orbital phase.  The marginal posteriors have been
  smoothed for better visualization.}
\label{fig:WASP43b_feng_retrieval}
\end{figure*}

\subsection{
Comparison to \citet{StevensonEtal2017ajWASP43bSpitzerPhase}
and \citet{MendoncaEtal2018ajWASP43bPhase}}
\label{sec:stevenson}

To compare our results with those shown in Fig.\ 6
of \citet{StevensonEtal2017ajWASP43bSpitzerPhase}, we retrieved both
the {\disk} and {\resolved} spectra at three orbital phases: 0.25
(first quarter or Eastern terminator), 0.5 (dayside), and 0.75 (third
quarter or Western
terminator). Figures \ref{fig:WASP43b_stevenson_spectra}
and \ref{fig:WASP43b_stevenson_composition} shows our retrieval
results.

Our retrieval analysis of the {\disk} dataset is generally consistent
with that of \citet{StevensonEtal2017ajWASP43bSpitzerPhase}.  In all
cases, the {\pyratbay} models fit the observations well, yielding
reduced chi-square values of $\chi_{\rm red}^2=1.0-1.5$.  We constrain
{\xwater} at all three phases, obtaining higher abundances for the
dayside spectrum.  At first and third quarters, {\water} shows a main
solution mode in the $\ttt{-5} < \xwater < \ttt{-4}$ range.  For this
mode, the mixing ratios of the three other molecules are not strongly
correlated with {\xwater}, and show mostly low values (see
Fig.~\ref{fig:WASP43b_stevenson_composition}).  Additionally, the
{\xwater} posterior shows a tail of higher values that are strongly
correlated with {\xcarbmono}, {\xcarbdiox}, and {\xmethane}, and
anti-correlated with {\pcloud}.  The strongest {\methane} absorption
band is probed by the {\Spitzer} 3.6-{\micron} filter, thus, this data
point (relative to the {\HST} observations) mainly determines the
{\xmethane} posterior distribution.  The {\xmethane} posteriors show
mostly upper-limit constraints, being more stringent at the dayside
phase ($\xmethane \lesssim 6$), and allowing for higher values at the
first- and third-quarter phases ($\xmethane \lesssim 4$).  In
contrast, \citet{StevensonEtal2017ajWASP43bSpitzerPhase} find tighter
constraints at $\xmethane\approx\ttt{-4}$.  This discrepancy is not
surprising, since we adopted the {\Spitzer} data reanalyzed
by \citet{MendoncaEtal2018ajWASP43bPhase}, which exhibits higher
planetary emission at 3.6~{\microns} (which contains the strongest
{\methane} band probed by these observations)
than \citet{StevensonEtal2017ajWASP43bSpitzerPhase}.  For a
non-inverted temperature profile, a stronger planet-to-star flux ratio
implies that there is less atmospheric absorption such that the
observations probe deeper, hotter layers.  To reduce the absorption,
the model thus requires a lower concentration of the dominant
absorber, in this case {\methane}.

Analogously to {\methane}, the strongest absorption bands for CO and
{\carbdiox} are probed by the {\Spitzer} 4.5-{\micron} channel.  This
broadband datum cannot distinguish CO from {\carbdiox} absorption,
leading to anti-correlated {\xcarbmono}--{\xcarbdiox} posteriors.
Compare for example the spectra at first and third quarters, since the
4.5~{\micron} eclipse depth at phase 0.75 is much lower than at phase
0.5, the data drive the posterior distribution towards higher
{\xcarbmono} or {\xcarbdiox} values (higher abundances increase the
absorption at high altitudes, where it is colder for a non-inverted
temperature profile, and hence produce lower planetary emission).  At
the dayside phase, {\xcarbmono/\xcarbdiox} posteriors prefer large
values to overcome the {\water} absorption at 4.5-{\micron}
absorption.  The pair-wise posteriors favor runs with a certain
minimum amount of {\xcarbmono}, {\xcarbdiox}, or both (right-bottom
set of panels in Fig.~\ref{fig:WASP43b_stevenson_composition}).

Our temperature posteriors are consistent with those
of \citet{StevensonEtal2017ajWASP43bSpitzerPhase} as well, showing
non-inverted temperature profiles in the $\sim$1000--1800~K range,
with photospheres concentrated near 10--$\ttt{-3}$ bar.  Finally, we
ran cloud-free retrievals to study the impact of the cloud model.  The
models are able to fit the data equally well with or without the gray
cloud model. Both cloudy and cloud-free modeling yield similar results
in terms of chemistry and temperature. The Bayesian information
criterion \citep[BIC,][]{Schwarz1978anstaBIC} therefore favors the
simpler cloud-free model.

\subsection{Longitudinally Resolved Retrieval Results}

We found that the {\resolved} spectra at first and third quarters are
practically indistinguishable from the {\disk} spectra (the
differences are smaller than the uncertainties,
Figs.\ \ref{fig:WASP43b_stevenson_spectra}
and \ref{fig:WASP43b_stevenson_composition}).  Consequently, the
{\resolved} retrieved results are nearly identical to those of the
{\disk} spectra.

For the dayside phase, the {\resolved} spectrum is brighter than the
{\disk} spectrum at all wavelengths, which is expected, since this
phase is located near the peak of the phase curve and the conversion
to {\resolved} spectra acts as a low pass filter.  Consequently, the
retrieval analysis found a temperature profile that remains hotter at
higher altitudes compared to the {\disk} profile
(Fig.~\ref{fig:WASP43b_stevenson_spectra}, middle panels).  In terms
of composition, the {\xcarbmono}, {\xcarbdiox}, and {\xmethane}
posterior distributions are consistent with the {\disk} posteriors.
Only the {\xwater} posterior is noticeably shifted to higher values,
although the medians of the {\resolved} and {\disk} posteriors are not
statistically different (Table \ref{table:WASP43b_retrieval}).

These results suggest that, given the available {\HST} and {\Spitzer}
phase-curve observations of WASP-43b, the {\resolved} analysis allows
us to detect a hotter temperature of the planet around its substellar
point (day-side phase), than that inferred from a {\disk} analysis.
However, the data do not have sufficient signal-to-noise to detect
spectral variations.  Therefore, the atmospheric composition estimated
from the {\resolved} analysis does not differ significantly from that
of the {\disk} analysis.

\subsection{Comparison to \citet{FengEtal2020aj2Dretrievals}}

To put the results of the {\resolved} retrieval approach in context
with other multi-dimensional retrieval approaches, we compared our the
WASP-43b retrieval analysis to those
of \citet{FengEtal2020aj2Dretrievals}.

We retrieved the WASP-43b phase-curve spectra at the 15 orbital phases
reported by \citet{StevensonEtal2017ajWASP43bSpitzerPhase} and, in
contrast to our previous analysis, we fit
the \citet{StevensonEtal2017ajWASP43bSpitzerPhase} for both {\HST} and
{\Spitzer} rather than fitting
the \citet{MendoncaEtal2018ajWASP43bPhase} {\Spitzer} data.  We
adopted the same system properties and atmospheric parameterization
of \citet{FengEtal2020aj2Dretrievals}.  For the temperature profile we
employed the Eddington approximation temperature profile with all five
parameters free, using their priors.  For the composition, we assumed
a cloud-free atmosphere and fit for the {\water}, CO, {\carbdiox}, and
{\methane} volume mixing ratios (neglecting {\ammonia} since it does
not impact the observed spectra).

Figure \ref{fig:WASP43b_feng_retrieval} shows our retrieval results.
In terms of the 1D {\disk} analysis, our composition posterior
distributions follow the same trends with orbital phase for all four
molecules as those of \citet{FengEtal2020aj2Dretrievals}.  The low
flux at the night side leads to unconstrained composition for all
molecules for the 2--4 phases near the anti-stellar point.  The
{\water} composition is constrained at all other phases at values
$\xwater\sim$$\ttt{-4}$, and generally increasing toward the hot-spot
longitude (phase $\sim$0.45). {\xcarbmono} is unconstrained, whereas
{\xcarbdiox} is largely unconstrained except at some phases after
secondary eclipse.  For {\methane} we found well-constrained
abundances around both quadratures at $\xmethane\sim\ttt{-5}$ and
upper limits around the hot-spot longitude.

Regarding the {\resolved} analysis, the retrieved compositions
remained largely consistent with those of the {\disk} analysis.  The
{\resolved} {\xwater} posteriors have somewhat higher values at
certain phases \citep[similar to][]{FengEtal2020aj2Dretrievals} and
the {\xmethane} posteriors show only modest differences with respect
the {\disk} posteriors (three well-constrained posteriors turned to
upper limits).  In comparison, the 2TP approach
of \citet{FengEtal2020aj2Dretrievals} found more distinct differences
in the {\methane} posteriors.  Out of the eight orbital phases with
well-constrained $\xmethane$ by their 1D model, the 2TP approach
favored upper limits at four phases.  At the remaining phases, the 2TP
$\xmethane$ posteriors peaked near their 1D counterparts, although
with lower precision \citep[see Fig.~6
of][]{FengEtal2020aj2Dretrievals}.

Conceptually, the temperature profiles
of \citet{FengEtal2020aj2Dretrievals} and our approaches are widely
different; while ours represents the local profile at a specific
longitude on the planet, theirs represents the profiles of hot and
cold components.  Thus, we cannot make a direct comparison between
these approaches.  However, generally, the two approaches obtained
similar non-inverted profiles along the orbit, ranging from $\sim$500
K to 1500 K between the night side and day side.

While the study of \citet{FengEtal2020aj2Dretrievals} indicates that
the 2TP approach can effectively recover more accurate physical
properties under the assumption of a two-component atmospheric model,
it is not possible to assess which approach is more accurate without
knowing a ground truth.  In the next section we aim to characterize
the validity of the {\resolved} approach by retrieving synthetic phase
curve spectra of atmospheric models that are more sophisticated than
in previous studies found in the literature.

\begin{figure*}[t]
\includegraphics[width=\linewidth, clip]{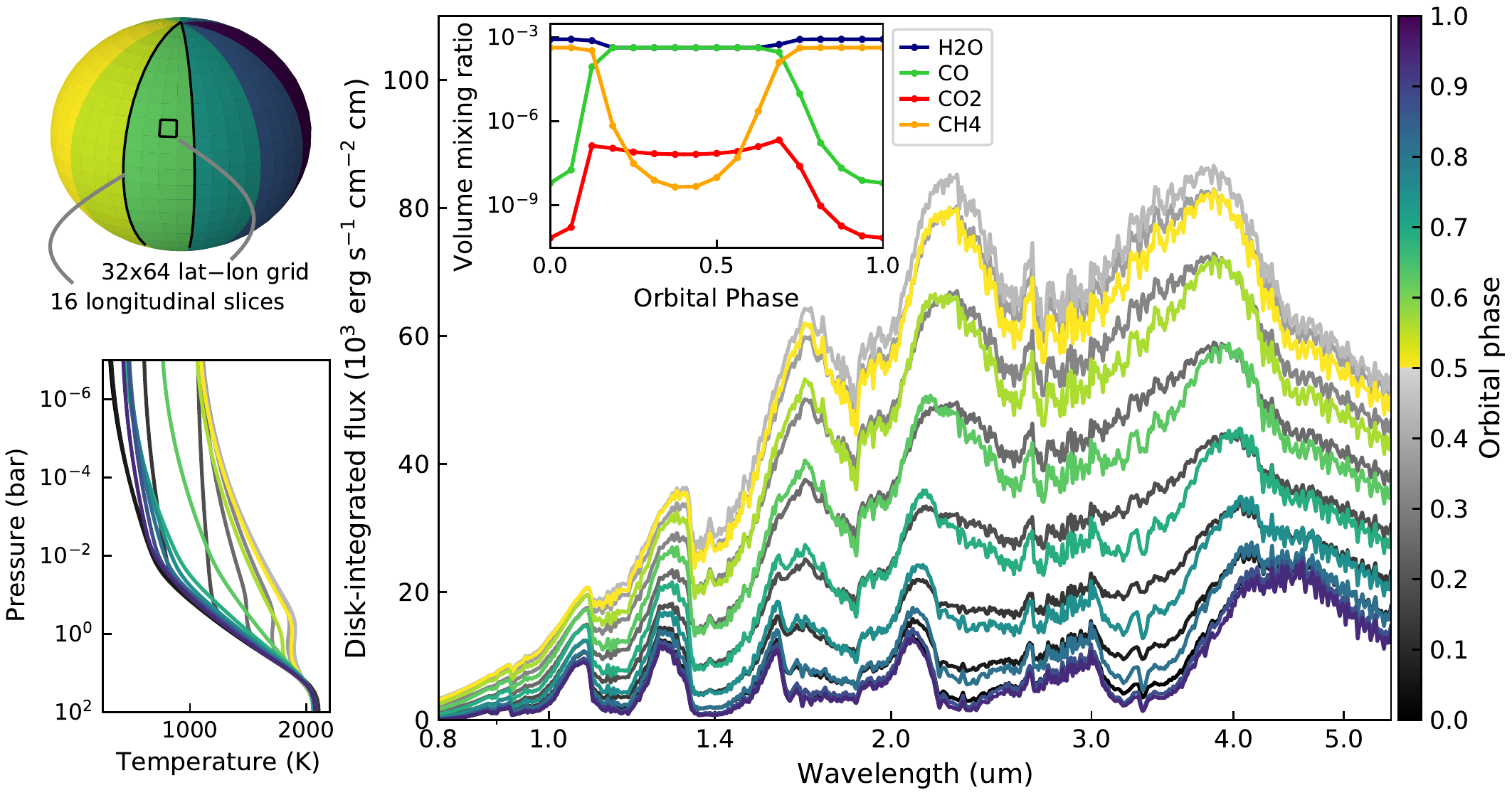}
\caption{
  Synthetic WASP-43b model. The sketch on the top left depicts the
  partitioning of the planetary model (16 longitudinal slices, each
  with a uniform temperature profile and composition).  The bottom
  left panel shows the temperature profile of the 16 slices color
  coded by orbital phase (color bar on the far right).  The right
  panel shows the {\disk} emission spectra as observed at 16 phases
  along the orbit.  The inset shows the volume mixing ratios of the
  spectroscopically active molecules.}
\label{fig:model_spectra}
\end{figure*}

\section{Retrieval of Synthetic {\JWST} WASP-43b Phase-curve Spectra}
\label{sec:synthetic}

We hypothesize that observations with greater constraining power than
the {\HST} and {\Spitzer} observations of WASP-43b would benefit more
from the use of {\resolved} spectral retrieval.  To test this
hypothesis, we performed atmospheric retrievals of synthetic WASP-43b
phase-curve observations with the upcoming {\Webb} ({\JWST}).

\subsection{Two Inhomogeneous Models of WASP-43b}

We based the inhomogeneous planetary models on the WASP-43b 3D global
circulation model (GCM)
of \citet{VenotEtal2020apjWASP43bWebbPredictions}, which we analyzed
with both the {\disk} and {\resolved} approaches.  The GCM solves a
coupled system of radiative-transfer and primitive equations on a
cube-sphere grid, considering the incident stellar irradiation of
WASP-43 and equilibrium molecular abundances for a cloudless
solar-composition model.  The uneven stellar forcing produces a large
day--night temperature contrast of $\sim$600--800~K at photospheric
pressures ($\sim$1~bar to 1~mbar).  Strong zonal winds lead to an
equatorial superrotation regime and a hotspot shifted eastward of the
sub-stellar point.  Thus, phase-curve observations generated from this
GCM have a peak infrared flux that occurs before secondary
eclipse \citep[see also][]{KatariaEtal2015apjWASP43bGCM}.
  
The GCM consists of a 3D grid of temperatures in terms of latitude
($\theta$), longitude ($\phi$), and pressure ($p$).  The latitude and
longitude grids are linearly spaced over the planet with 32 and 64
samples, respectively.  The pressure profile is logarithmically spaced
between 170~bar and 3$\tttt{-6}$~bar with 52 samples.  The longitude
origin is located at the sub-stellar point and increases eastward.  In
this scheme, the planet rotates along the Z axis; the orbital phase
increases from $0.0$ (nightside facing the observer, transit), to
$0.5$ (dayside facing the observer, eclipse), to $1.0$ (nightside
facing the observer, transit).  Note that in the Appendices the
orbital phase is measured in radians and with the origin at eclipse
midtime.

In our framework, we kept the 32$\times$64 latitude--longitude grid,
but to simplify the interpretation of the retrieval analysis, we
divided the planet into 16 longitudinal slices, where all grid cells
in a given slice have the same vertical temperature and composition
profiles (Figure \ref{fig:model_spectra}, top--left panel).  We
assumed a circular orbit with $90^\circ$ inclination and zero
obliquity.

We computed the temperature profile for each slice by selecting a set
of GCM profiles evenly spaced in longitude and at a latitude of
$30^\circ$S.  We fit these temperature profiles with
the \citet{MadhusudhanSeager2009apjRetrieval} parametric temperature
model.  This model divides the atmosphere into three regions,
delimited by pressures $p_1$ and $p_3$:
\begin{equation}
T(p) = \left\{
\begin{array}{lll}
T_0 + \left[\frac{1}{a_1}\ln(p/p_0)\right]^2 & \text{if } p < p_1
   & (\rm layer\ 1) \\
T_2 + \left[\frac{1}{a_2}\ln(p/p_2)\right]^2 & \text{if } p_1 \le p < p_3
   & (\rm layer\ 2) \\
T_3   & \text{if } p \ge p_3 & (\rm layer\ 3)
\end{array} \right.
\end{equation}
with $T_0$ the temperature at the top of the atmosphere ($p_0$), and
parameters $a_1$, $a_2$, and $p_2$ controlling the variation of
temperature with pressure ($T_2$ and $T_3$ can be derived by
evaluating the model at the layers' boundaries).  During this fitting
step, we resampled the profile into our 100--$\ttt{-8}$~bar pressure
range from \S\ref{sec:wasp43_retrieval}, and turned small thermal
inversions near the 1-bar level into isothermal regions
(Fig.\ \ref{fig:model_spectra}, bottom--left panel).

We considered two levels of complexity for the atmospheric model
(Table \ref{table:WASP43b_synthetic_models}).  The simpler case (M1)
assumed a symmetric day-to-night temperature profile variation for the
16 longitudinal slices by adopting the 9 temperature profiles of the
evening side, folding them over the entire orbit symmetrically around
the hottest profile placed at secondary eclipse (see viridis
color-mapped profiles in Fig.\ \ref{fig:model_spectra}).  For the
composition we assumed globally constant abundances adopting the
values from \citet{FengEtal2020aj2Dretrievals}, i.e.:
$\xwater=4\tttt{-4}$, $\xcarbmono=2\tttt{-4}$, and
$\xcarbdiox=\xmethane=\ttt{-9}$.

\begin{table}[htb]
\centering
\caption{Synthetic WASP-43b Models}
\label{table:WASP43b_synthetic_models}
\begin{tabular*}{0.95\linewidth} {@{\extracolsep{\fill}} cll}
\hline
\hline
Model  & Temperature Profile & Composition  \\
  \hline
  M1 & day--night symmetric & globally constant \\
  M2 & asymmetric, hot-spot offset & varying with longitude \\
  \hline
\end{tabular*}
\end{table}

For the more complex case (M2), we adopted the 16 asymmetric
temperature profiles derived from the GCM
(Fig.\ \ref{fig:model_spectra}).  We adopted a cloud-free solar
elemental composition with vertically uniform volume mixing ratios but
varying with longitude.  Although the vertically uniform assumption
clearly deviates from self-consistent chemical or radiative
prescriptions, at present this is a common assumption for exoplanets.
The use of more sophisticated models is impractical for exoplanet
retrievals since they show a higher complexity than what can be
constrained by observations and have higher computational demands than
simpler parametric models.  Furthermore, the limited spectral range of
the observing instruments limits the pressure range of the atmosphere
being probed.  We considered the opacity of {\molhyd} and He (dominant
gases), and {\water}, CO, {\carbdiox}, and {\methane} (trace gases).
To compute the composition at each longitudinal slice, we calculated
thermochemical-equilibrium abundances using the open-source
code \textsc{RATE} \citep{CubillosEtal2019apjRate}, evaluated at the
temperatures near the 0.1-bar level.  This choice leads to a
well-defined transition between {\methane} and CO/{\carbdiox} trace
gases between the night side and the day side
(Fig.\ \ref{fig:model_spectra}, inset).  We did not take into account
reflected light, which could be significant in the near
infrared \citep{KeatingCowan2017apjWASP43bHeatTransport}.

To generate the phase-curve emission spectra we followed the procedure
of \citet{BlecicEtal2017apj3Dretrieval}.  We computed the {\disk}
spectra as the sum of the intensities from each latitude--longitude
cell, weighted by their projected area in the direction of the
observer:
\begin{equation}
\label{eq:disk_flux}
F_{\lambda}(\xi) = \sum_{\theta,\phi}
    I_{\lambda}(\theta,\phi,\xi)\ \mu(\theta,\phi,\xi)\ \Delta A(\theta,\phi),
\end{equation}
where, for each cell, the surface element is $\Delta A(\theta,\phi)
= \sin\theta \Delta\theta \Delta\phi$, and the angle from the normal
vector is $\mu(\theta,\phi,\xi) = \sin\theta \cos(\phi-\phi_\xi)$,
with $\phi_\xi$ the sub-observer longitude at orbital phase $\xi$.
Note that Eq.\ (\ref{eq:disk_flux}) sums only over the cells in the
hemisphere visible to the observer, i.e., $\mu>0$.  Under the
local-thermodynamic-equilibrium and plane-parallel approximations, the
emergent intensity spectrum is given by
\begin{equation}
I_{\lambda}(\theta,\phi,\xi) = \int_{0}^{\tau_{\rm max}}
   \frac{1}{\mu} B_{\lambda}(\tau) e^{-\tau/\mu} {\rm d}\tau,
\end{equation}
where $B_{\lambda}$ is the Planck function, and $\tau=\tau_{\lambda}$
is the optical depth, which is integrated from the top of the
atmosphere ($\tau_{\lambda}=0$) to the bottom
($\tau_{\lambda}=\tau_{\rm max}$).  Figure \ref{fig:model_spectra}
(right panel) shows the WASP-43b {\disk} model spectra evaluated at 16
phases along the orbit.  These models consider the same wavelength
sampling and opacities sources as in \S\ref{sec:wasp43_retrieval}, but
range from 0.8 to 5.5~{\microns}.

Note that since the underlying GCM input has an equatorial jet that
transports energy eastward \citep{KatariaEtal2015apjWASP43bGCM}, the
phase curve is characterized by a hot-spot offset (peak emission
occurs near phase 0.44, slightly before secondary eclipse) and an
asymmetric variation (the observed spectrum `heats up' faster than it
`cools down' around the hot spot).  This asymmetry has clear
repercussions in the composition, for example, the methane feature at
3.4~{\microns} (Fig.~\ref{fig:model_spectra}, right panel) is more
prominent at the evening phases (0.5--1.0) than at the morning phases
(0.0--0.5).  Our setup is distinctly more complex than other recent
multi-dimensional atmospheric retrieval studies, which only vary the
temperature profile with longitude while keeping a globally
homogeneous composition; however, we believe that this is necessary to
better capture the complexity of exoplanet atmospheres, and to prevent
arriving at overly optimistic results.

\subsection{Simulated JWST Observations}

We modeled expected {\JWST} performance using the {\pandexo}
interface \citep{BatalhaEtal2017paspPandexo} to the {\JWST} Exposure
Time Calculator\footnote{\href{https://jwst.etc.stsci.edu/}
{https://jwst.etc.stsci.edu/}} \citep{PontoppidanEtal2016spiePandeia}.
We selected the NIRISS SOSS and NIRSpec G395H instruments since they
provide a nearly complete coverage from 0.8 to 5.2 {\microns} with
only two instruments, and better signal-to-noise ratios than
alternative instruments (NIRSpec PRISM could cover the entire 0.8--5.0
{\micron} range in a single observation but with notably lower
signal-to-noise ratios).  We followed the same assumptions
as \citet{VenotEtal2020apjWASP43bWebbPredictions} to estimate the
observing times, i.e., we observe each phase for 1/16th of the orbital
period with a baseline of twice the eclipse duration.  We let
{\pandexo} optimize the exposure parameters, reaching 80\% of the
detector's maximum saturation level.  We assumed a stellar K-band
magnitude of 9.27, based on the Kurucz model of the star.

For NIRISS SOSS, we selected the GR700XD disperser, SUBSTRIP96
subarray and NISRAPID readout pattern (0.8--2.8~{\microns} range),
with a noise floor of 20 ppm, following the expected instrumental
noise level \citep{GreeneEtal2016apjJWSTtransitCharacterization}.
This yields 113 integrations of 38 groups for each orbital phase,
leading to a SNR of 5900 at 1.5~{\microns}.  For NIRSpec, we selected
the G395H grating, S1600A1 fixed slit, SUB2048 subarray, and NRSRAPID
readout pattern (2.9--3.7~{\microns} and 3.8--5.2~{\microns} ranges),
with a noise floor of 30 ppm, as well as the aperture spectral
extraction strategy, with a target aperture of 0.75 arcsec, and a sky
band ranging from 0.75 to 1.5 arcsec.  This yields 95 integrations of
18 groups for each orbital phase, leading to a SNR of 430 at
4~{\microns}.  The {\JWST} Exposure Time Calculator reported no
warnings nor errors for any of these calculations.

We adopted as the {\disk} dataset the model described in the previous
section, which we binned to a resolving power of $R \approx 200$ and
propagated the {\pandexo} uncertainties accordingly.  We did not add
random noise to the data points \citep[following suite
with,][]{FengEtal2018apjEarthAnalogReflectedRetrievals,
MaiLine2019apjTransmissionCloudsJWST,
TaylorEtal2020mnrasEmissionBiases}, such that we can attribute any
inaccuracy of the retrieval to the methodology rather than to
statistical random sampling from the true model.
A side effect of this choice is that we no longer expect the
$\chi_{\rm red}^2$ to approach unity.  We constructed the {\resolved}
datasets by fitting a Fourier series to the simulated {\disk} spectra
and repeating the procedure of \S\ref{sec:wasp43_retrieval}.

To determine the appropriate order of Fourier series for fitting the
simulated data, we fit several series of sinusoids to the simulated
WASP-43b phase curves at each wavelength, from first order to fourth
order. We used PyMC3 \citep{SalvatierEtal2016asclPyMC3} to fit the
data and estimate uncertainties. For each model, we calculated the
Widely Applicable Information
Criteria \citep[WAIC;][]{Watanabe2010arxivWAIC} for each model. The
WAIC is similar to the BIC, but uses the full fit posteriors. The
second order model was favoured over the other models, meaning we can
detect first and second order modes but are insensitive to higher
orders. At each step of the MCMC, we calculated the
wavelength-dependent longitudinal brightness maps and associated
uncertainties.

We analyzed the synthetic data in the same manner as the
{\HST}+{\Spitzer} observations, except that in this case we assumed a
cloud-free atmosphere and used
the \citet{MadhusudhanSeager2009apjRetrieval} temperature-profile
parameterization (same model parameterization as that used to
construct the input model).
Table \ref{table:WASP43b_synthetic_retrieval} summarizes the model
parameterization for the synthetic {\JWST} WASP-43b retrievals.

\begin{table}[tb]
\centering
\caption{Synthetic {\JWST} WASP-43b Retrieval Parameterization}
\label{table:WASP43b_synthetic_retrieval}
\begin{tabular*}{0.8\linewidth} {@{\extracolsep{\fill}} lll}
\hline
\hline
Parameter  & Priors  \\
\hline
$\log_{10}(p_1/{\rm bar})$ & $\mathcal U(-7, 2)$  \\
$\log_{10}(p_2/{\rm bar})$ & $\mathcal U(-7, 2)$  \\
$\log_{10}(p_3/{\rm bar})$ & $\mathcal U(-5, 2)$ \quad and \quad $p_3 > p_1$ \\
$a_1$ (K$^{-0.5}$)         & $\mathcal U(0, 2)$   \\
$a_2$ (K$^{-0.5}$)         & $\mathcal U(0, 2)$   \\
$T_0$ (K)               & $\mathcal U(100, 3000)$  \\
$\log_{10}(X_{\rm H2O})$  & $\mathcal U(-12, -1)$    \\
$\log_{10}(X_{\rm CO})$   & $\mathcal U(-12, -1)$    \\
$\log_{10}(X_{\rm CO2})$  & $\mathcal U(-12, -1)$    \\
$\log_{10}(X_{\rm CH4})$  & $\mathcal U(-12, -1)$    \\
\hline
\end{tabular*}
\end{table}

\begin{figure*}[t]
\centering
\includegraphics[width=\linewidth, clip]{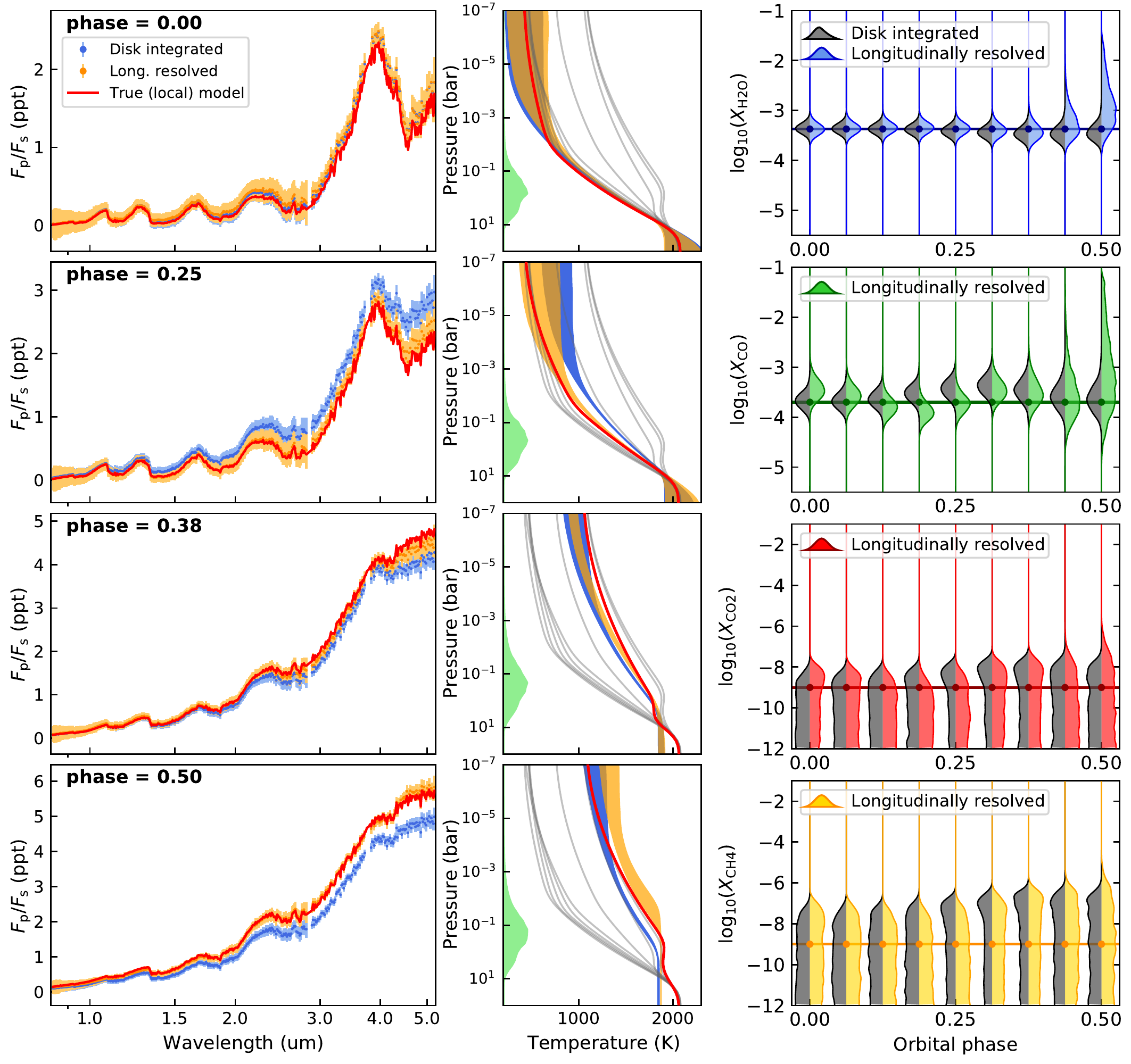}
\caption{Model and retrieval of the synthetic WASP-43b {\JWST}
    phase-curve observation of the constant-composition case (M1).  {\bf
      Left panels:} the markers with error bars denote the {\disk}
    (blue) and {\resolved} (orange) planet-to-star flux ratios at
    selected phases (from top to bottom, see labels).  The red curves
    denote the spectra generated from a 1D model using the local
    conditions of the true model at the sub-observer longitude of each
    orbital phase.  {\bf Center panels:} retrieved temperature
    profiles of the synthetic datasets from the left panel.  The blue
    and orange shaded areas denote the {\disk} and {\resolved}
    temperature-profile posterior distributions, respectively (central
    68\% credible interval of the posteriors).  The red curves denote
    the temperature profile at the sub-observer longitude of each
    phase. The gray curves denote the temperature profiles of the true
    model at all other longitudinal slices.  The width of the green
    curve denotes the relative contribution function at each pressure
    level for these {\JWST} simulations. {\bf Right panels:}
    volume-mixing-ratio posterior distributions for the {\disk} (gray)
    and {\resolved} (colored) analyses as a function of orbital phase
    (only shown between transit and eclipse epoch, since the model is
    symmetric).  The marginal posteriors have been smoothed for better
    visualization.  The solid curves with dot markers denote the
    abundances of the true model.}
\label{fig:model_retrieval1}
\end{figure*}

\subsection{
  Retrieval of Synthetic WASP-43b Phase Curve with Constant Composition (M1)}
\label{sec:retrieval_simulated1}

Figure \ref{fig:model_retrieval1} (left panels) show a sample of the
synthetic {\disk} and {\resolved} spectra.  This figure also shows 1D
model spectra assuming the local properties at the corresponding
sub-observer longitude of each phase.  Unlike the {\HST}/{\Spitzer}
spectra, the simulated {\JWST} datasets show significant differences
between the {\disk} and {\resolved} spectra, which are both phase- and
wavelength-dependent.  At all phases the {\resolved} spectra reproduce
better the local spectra of the respective sub-observer longitudes,
particularly near quadrature (phases 0.25) and secondary-eclipse epoch
(phase 0.5).

The retrieved temperature profiles follow the same trend as the
spectra, with the {\resolved} profiles matching the local profiles
better than the {\disk} profiles.  Over the $10$--$\ttt{-3}$~bar
region probed by the observations the {\resolved} retrieval always
reproduces the local temperature of the planet, whereas the {\disk}
retrieval can miss the true temperatures by up to $\sim$100~K
(many times the credible intervals).  The magnitude of this mismatch
is consistent with the mismatch in the spectra at the respective
orbital phases.

In terms of composition, the {\disk} analysis retrieves the true
values within the posterior credible intervals (constraining the
{\water} and CO volume-mixing ratios and finding upper limits for
{\carbdiox} and {\methane}).  The {\resolved} analysis produces
similar results, well in agreement with the true values as well.

\begin{figure*}[t]
\centering
\includegraphics[width=\linewidth, clip]{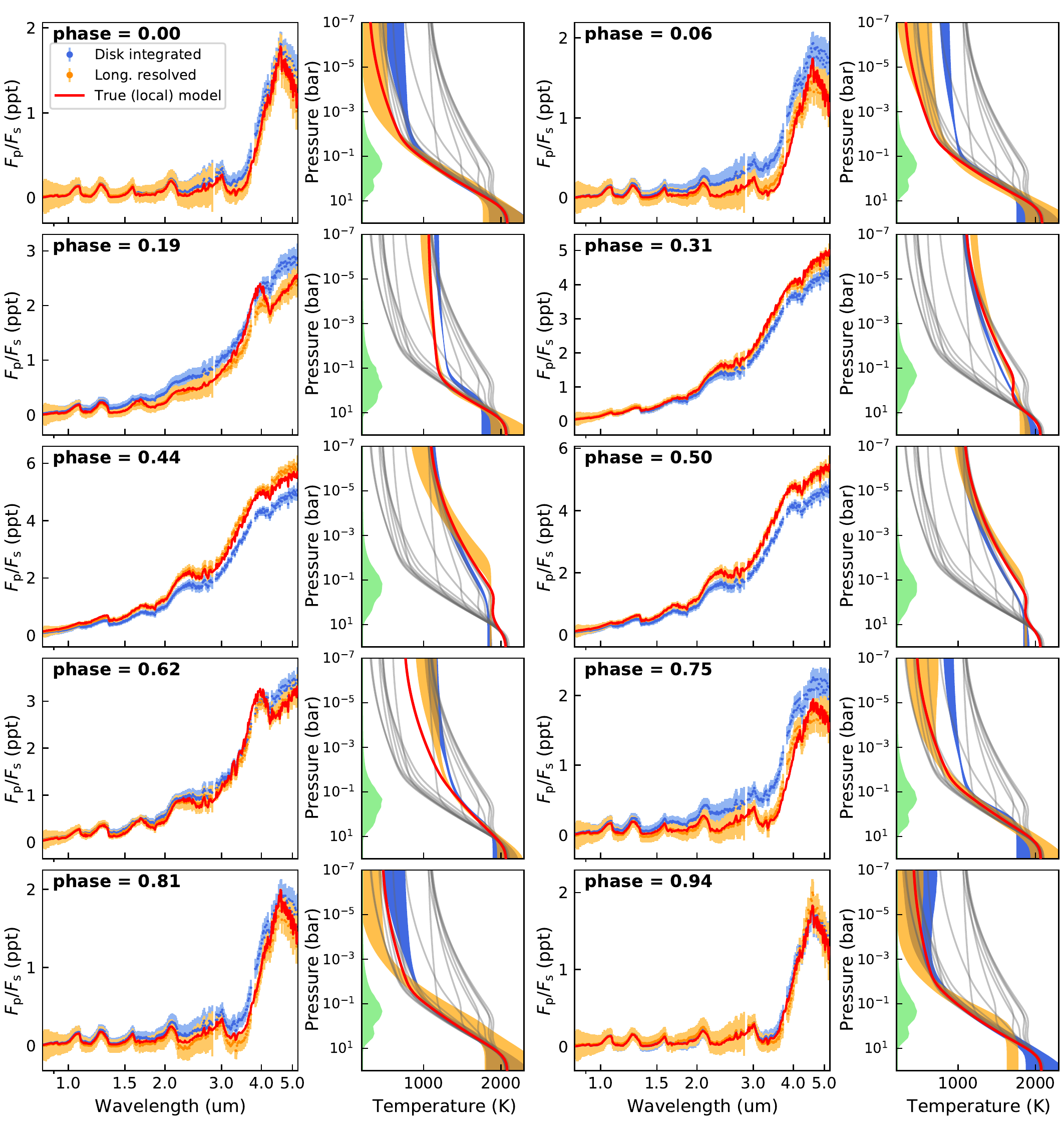}
\caption{
  Model and retrieval of the synthetic WASP-43b {\JWST} phase-curve
  observation of the variable-composition case (M2).  Each pair of
  panels show the model spectra and retrieved temperature profiles at
  selected orbital phases (see labels).  The {\disk} and {\resolved}
  analyses are denoted by the blue and orange colors, respectively.
  The red curves denote local spectra and temperature profiles at the
  sub-observer longitude of each orbital phase.  The gray curves
  denote the temperature profiles of the true model at all
  longitudinal slices.  The width of the green curve denotes the
  relative contribution function at each pressure level for these
  {\JWST} simulations.}
\label{fig:model_retrieval2_spectra}
\end{figure*}

\begin{figure}[t]
\includegraphics[width=\linewidth,clip]{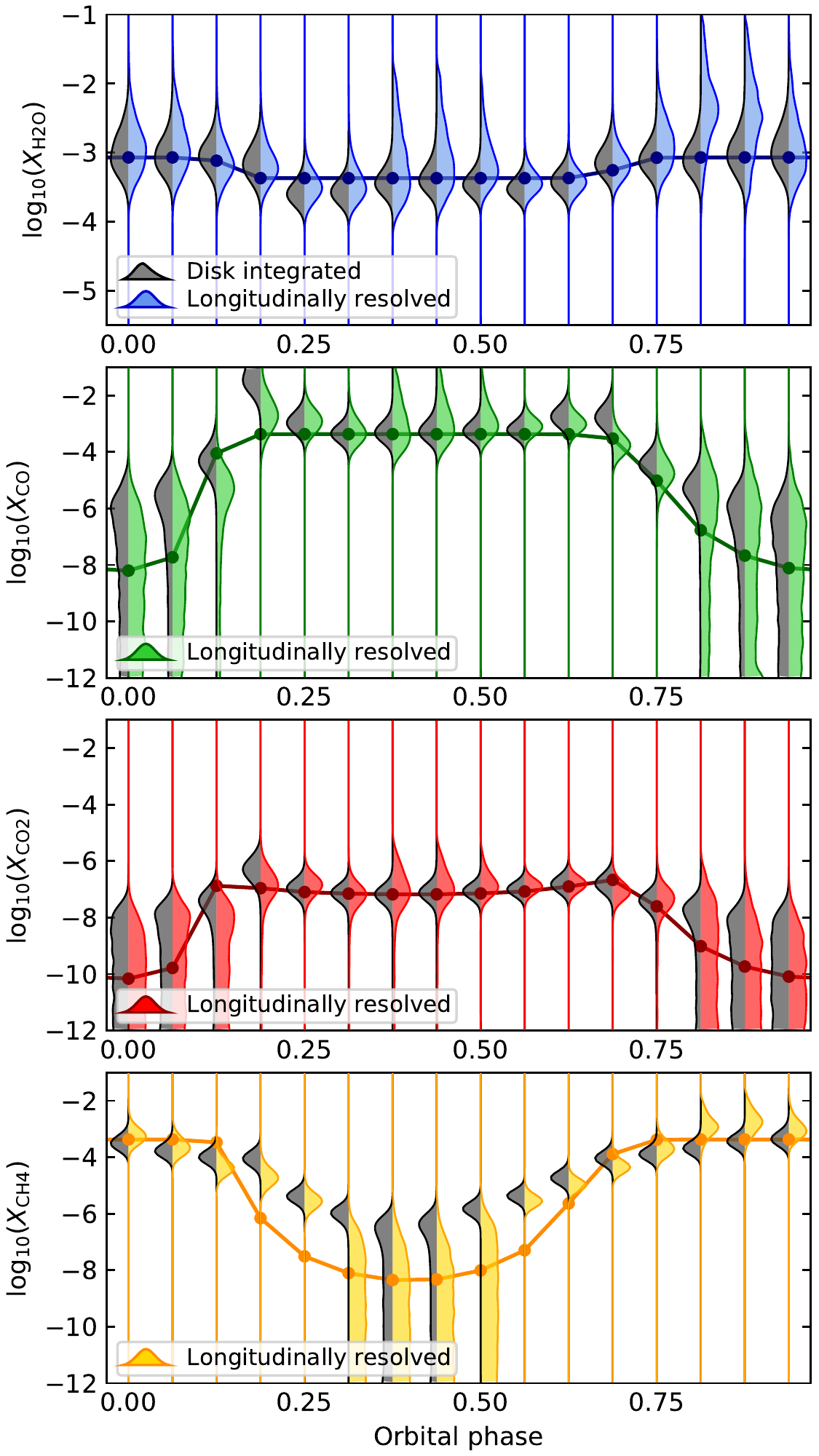}
\caption{
  Retrieved WASP-43b volume-mixing ratios of the synthetic WASP-43b
  {\JWST} phase-curve observation of the variable-composition case
  (M2).  The gray and colored histograms show the posterior
  distributions of the {\disk} and {\resolved} analyses, respectively,
  as a function of orbital phase.  The marginal posteriors have been
  smoothed for better visualization.  The solid curves with dot
  markers denote the input abundances at the sub-observer longitudes
  corresponding to each orbital phase.}
\label{fig:model_retrieval2_abundances}
\end{figure}

\subsection{
    Retrieval of Synthetic WASP-43b Phase Curve with Variable Composition (M2)}
\label{sec:retrieval_simulated2}

Figure \ref{fig:model_retrieval2_spectra} (left-side panels) shows a
sample of the synthetic {\disk} (blue), {\resolved} (orange), and true
model at the local sub-observer longitude (red) spectra.  Once again,
the simulated {\JWST} datasets show significant differences between
the {\disk} and {\resolved} spectra.

Both the {\disk} and {\resolved} retrievals fit the data (not shown);
however, the {\resolved} spectra give a more accurate representation
of the local spectra.  Overall, the spectra are more uniform on the
night side than on the day side: the {\disk} and {\resolved} spectra
are consistent with each other at the longitude opposite the hot-spot
(phase = 0.94).  At the surrounding longitudes, the {\disk} spectra
overestimate the local planetary emission because of the contribution
from the hotter dayside flux contributing to the observed hemisphere.
Around the hot-spot longitude we have the opposite situation, the
{\disk} spectra underestimate the local planetary emission since the
hemispheric integration dampens the signal.  The {\resolved} spectra
provide a much better match to the local spectra at all phases.

Figure \ref{fig:model_retrieval2_spectra} (right-side panels) shows
the retrieved temperature-profile posteriors at selected orbital
phases.  The simulated observations probe mainly between $\ttt{-3}$
and 10~bar as shown by the contribution functions (green curves), and
thus the upper and lower end of the profiles are an extrapolation of
the parametric profiles.  We found that the {\resolved} temperature
posteriors are much more consistent with the true temperature profiles
than the {\disk} posteriors at most orbital phases (the {\resolved}
fits are sometimes on par with the {\disk}, but never worse).

Figure \ref{fig:model_retrieval2_abundances} shows the retrieved
volume-mixing ratios as a function of orbital phase.  Each of the four
molecules is spectroscopically detectable at least during some orbital
phases, though not all simultaneously.  Given the assumed
pseudo-thermochemical equilibrium chemistry of the input model,
{\methane} is abundant and hence detectable at the cooler night-side
longitudes at the expense of CO and {\carbdiox}; as one approaches the
hot spot, {\methane} turns into CO and {\carbdiox} due to the higher
temperature, making CO and {\carbdiox} detectable at the dayside
longitudes.  As expected, the {\xwater} constraint was the most
accurate and precise of the four molecules.  The {\resolved} post
processing seemed to have lowered the accuracy and precision at
specific phases, but the posteriors still capture the true value
within the 68\% credible intervals.  Possibly, the high abundance and
weak variation of {\xwater} along the orbit helps the retrievals to
fit this molecule more accurately.

In contrast, {\xcarbmono} and {\xcarbdiox} vary significantly along
the orbit.  At the dayside longitudes, {\xcarbmono} and {\xcarbdiox}
are well-constrained since CO and {\carbdiox} are abundant enough to
show detectable spectral features.  At the night-side longitudes, the
CO and {\carbdiox} abundances drop below detectable levels, leading to
upper-limit posteriors for {\xcarbmono} and {\xcarbdiox}.  The
{\resolved} {\xcarbmono} and {\xcarbdiox} posteriors are in good
agreement with the true local values at nearly all phases (i.e.,
within the 68\% credible interval).  The {\disk} {\xcarbmono} and
{\xcarbdiox} posteriors also trace the behavior of the true values,
but perform particularly poorly near first- and third-quarter phases.
A possible explanation could be that these are the phases where the
abundances vary the most across the visible hemisphere.  Considering
that the abundance estimations are strongly correlated between the
different species, and that the strength of the spectral features do
not scale linearly with the abundances, biases propagate particularly
strongly at these highly heterogeneous phases.

The {\methane} retrieval showed the poorest results.  At the
night-side longitudes, {\xmethane} remains abundant and constant.
Both {\disk} and {\resolved} {\xmethane} posteriors recover the
expected values; however, the true {\xmethane} values often fall
outside the posterior credible intervals.  As the orbital phase
approaches the hot-spot longitude, {\xmethane} drops below
spectroscopically detectable values.  Despite this significant
decrease, the {\disk} analysis shows tightly constrained {\xmethane}
posteriors that overestimate the true values by several orders of
magnitude.  This clear mismatch is reminiscent of the {\methane}
biases found by \citet{FengEtal2016apjNonUniformTempProfiles,
FengEtal2020aj2Dretrievals}.  The {\resolved} analysis effectively
recovered {\xmethane} upper limits at the four orbital phases around
the hot-spot longitude, but did not fit well the abundances around
first- and third-quarter phases.  This bias is significantly more
severe than for CO or {\carbdiox}.  In any case, it should be noted
that our simulated observations presume local-thermochemical
equilibrium, whereas horizontal quenching can drive the atmosphere
towards longitudinally uniform
abundances \citep{CooperShowman2008apjDynamicsAndChemistry}.

To quantify whether the {\resolved} retrievals improved the
volume-mixing ratio estimation over the {\disk} retrievals, we
computed the Z-scores between the true volume mixing ratios and the
marginal posterior distributions.  For a distribution with standard
deviation $\sigma$ and mean $\mu$, the Z-score of a datum $x$ is its
distance from the mean in units of the standard deviation,
$z=(x-\mu)/\sigma$.  Since the posteriors are often asymmetric and we
are interested in the maximum {\em a posteriori} estimation, we
computed the Z-scores using the mode of the posteriors.
Table \ref{table:retrieval_stats} presents the retrieval Z-score
(average of the Z-score absolute values over all orbital phases) for
each molecule and dataset.  The Z-scores show that the {\resolved}
analysis clearly improves the retrieval of {\xcarbmono} and
{\xmethane} (i.e., Z-scores closer to zero).  For the other two
molecules, both analyzes perform relatively well (Z-scores $<$ 1.0),
and the Z-score differences are small between the {\resolved} and
{\disk} analyses.

\begin{table}[h]
\centering
\caption{Retrieval Z-score Statistics}
\label{table:retrieval_stats}
\begin{tabular*}{\linewidth} {@{\extracolsep{\fill}} lccccl}
\hline
\hline
   & {\xwater} & \xcarbmono & \xcarbdiox & \xmethane & Approach \\
\hline
Z-score & 0.50 & 1.15 & 0.72 & 4.56 & disk{\hyph}integrated \\
Z-score & {\bf 0.41} & {\bf 0.69} & {\bf 0.53} & {\bf 2.76} & long. resolved \\
\hline
\end{tabular*}
\end{table}

\section{Discussion}
\label{sec:discussion}

\subsection{Longitudinally Resolved Spectral Retrieval: the Good, the
  Bad and the Ugly}
\label{sec:ugly}

We have shown that longitudinally resolved spectral retrieval is
preferable to {\disk} retrieval: when measurement uncertainties are
large the two approaches are equivalent, but for high signal-to-noise
observations it yields significantly different atmospheric properties
that are more accurate than {\disk} spectral retrieval. Moreover, our
approach is compatible with off-the-shelf 1D spectral retrieval models
and hence is complementary to efforts by other groups to develop
multi-dimensional spectral retrieval
codes \citep[][]{CaldasEtal2019aaTransmission3Deffects,
IrwinEtal2020mnras2.5Dretrieval, TaylorEtal2020mnrasEmissionBiases,
FengEtal2020aj2Dretrievals}.  While it may seem strange to perform
spectral retrieval on a reconstructed spectrum that was never directly
observed, this is conceptually similar to the data reduction and
decorrelation required for most exoplanet spectroscopy.

It should be emphasized that phase mapping is always an
under-constrained problem, regardless of the adopted parameterization.
The best one can do is quantify the degeneracies and how they impact
the astrophysical uncertainties.  For example,
{\cite{BeattyEtal2019ajKELT1bPhase}} suggested that certain phase
curves are more efficiently fit with a different map parameterization.
If the brightness map of a planet includes sharp transitions then a
map parameterization involving uniform longitudinal
slices \citep[cf.][]{KnutsonEtal2007nat189733Phase} may provide an
acceptable fit with fewer parameters than the Fourier mapping approach
adopted in this paper \citep[see updated discussion of mapping
parameterization in][]{CowanFujii2018bookMappingExoplanets}.

Marginalizing over different mapping parameterizations like sinusoids,
slices, or the step function from \cite{BeattyEtal2019ajKELT1bPhase}
would require a reversible jump MCMC. Alternatively, a pixel- or
slice-based map regularized with a Gaussian process can retrieve the
map and length scale of brightness
variations \citep{FarrEtal2018ajExocartographerReflectedLightMapping}.
However, any periodic function can be represented by a Fourier series,
it is just a question of the order of the expansion.

\begin{figure}[t]
\centering
\includegraphics[width=\linewidth, clip]{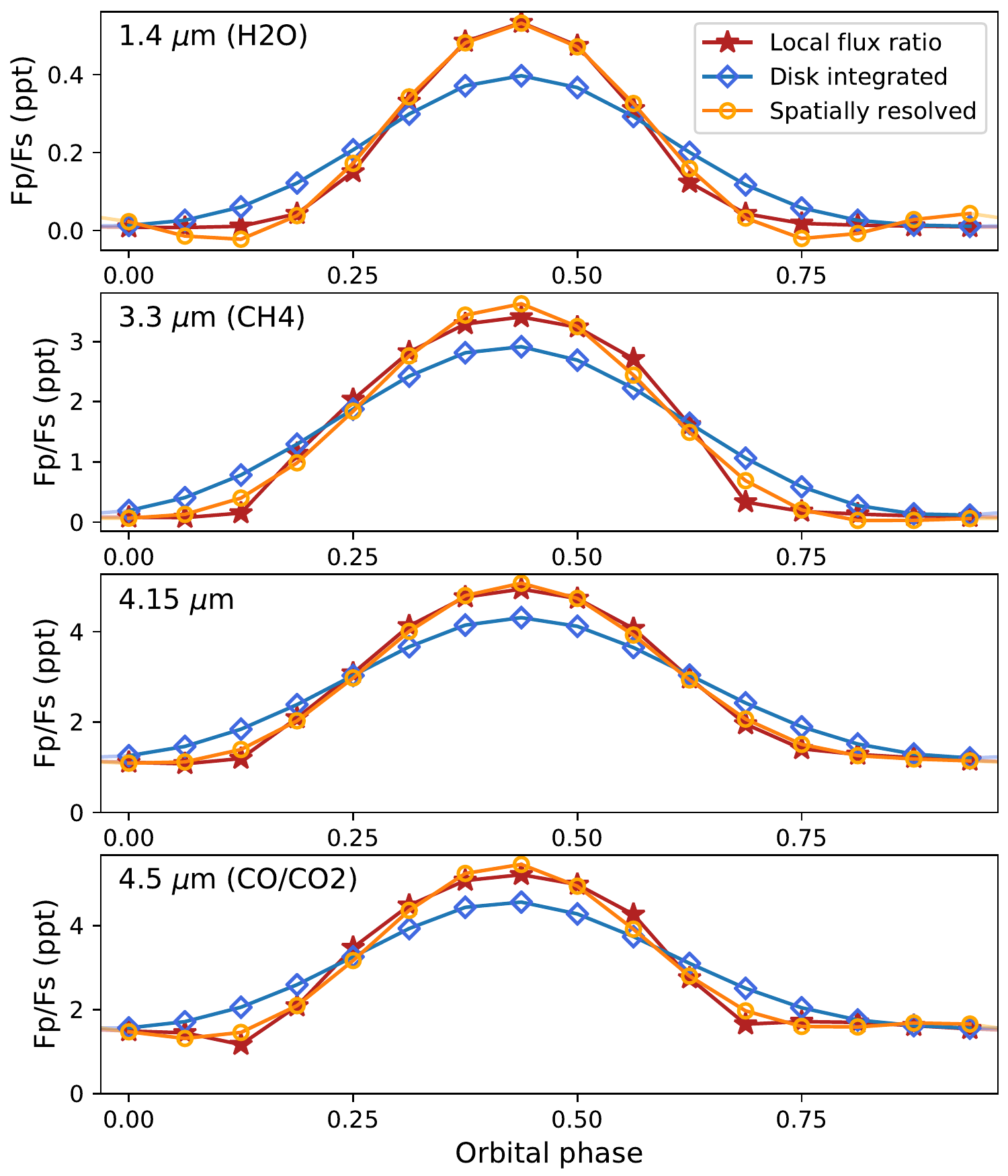}
\caption{
WASP-43b model phase curves at selected wavelengths, where the
spectrum is dominated by {\water} (1.4~{\micron}), {\methane}
(3.3~{\micron}), by no particular molecule (4.15~{\micron}), and
CO/{\carbdiox} (4.5~{\micron}).  The {\resolved} dataset matches much
better the phase curve assuming local properties at each phase than
the {\disk} dataset.  However, the spectral mapping is not perfect, in
particular, it does not perform well at first and third quadrature,
this is consistent with these phases being the most problematic for
the atmospheric retrievals.}
\label{fig:model_phasecurves}
\end{figure}

We have tested whether the data favour sharp discontinuities by
performing fits with different orders of Fourier series to our more
realistic simulated WASP-43b observations (M2 phase-curve model): maps
with sharp turns or discontinuities
(Figure \ref{fig:model_phasecurves}) produce light curves with
significant power in high-frequency Fourier modes.  The Information
Criteria suggest that second-order sinusoidal light curve models were
favoured over lower and higher order models.

For the second order Fourier series we adopt throughout this paper,
one can extract approximately four independent longitudinally resolved
spectra. Spectra at longitudes less than 90 degrees from each other
will be more or less correlated. This correlation is dictated by the
disk-integration rather than the specific map parameterization: a
pixel-based map with more than four slices would exhibit similar
correlations.

\subsection{Fourier Odd Harmonics}
\label{sec:odd}

For a planet on a circular edge-on orbit, symmetry dictates that odd
order harmonics present in the longitudinal brightness map are not
visible in the corresponding phase
curve \citep{CowanAgol2008apjPhaseInversion}.  Significant North-South
asymmetry in the temperature map could make odd harmonics visible in
the phase variations of a non-edge-on
planet \citep{CowanEtal2017mnrasPhaseOddHarmonics}, but most odd modes
of the map have no lightcurve signature, regardless of
inclination \citep{CowanEtal2013mnrasLightCurves}.  The odd-mode
degeneracy affects all map parameterizations: the addition of odd
longitudinal modes to any brightness map will not change the resulting
disk-integrated phase variations.

Since we simulated the WASP-43b observations using results from a GCM,
we know what the ``true'' longitudinal maps look like.  To quantify
how much the odd mode degeneracy could affect spectral mapping, we fit
the longitudinal maps from the WASP-43b GCM results directly, and
tested various order Fourier series like we did for the phase curves
(as in \S\ref{sec:ugly}).  This time, a third order model was favoured
over a second order model.  The first and second order amplitudes and
offsets were the same between the two models, because sinusoids of
various orders form an orthogonal basis.  Averaged over wavelengths,
the third order amplitude was just 9\% of the first order amplitude
for the simulated data.  Both the second-order and third-order maps
provide good fits to the true maps, matching within the map
uncertainties.  This suggests that, for a realistic GCM, the third
order degeneracy is not introducing an appreciable bias.

\subsection{When is Longitudinally Resolved Spectral Retrieval Preferable to Disk-Integrated Retrieval?}

The numerical experiments in \S\ref{sec:retrieval_simulated1}
and \S\ref{sec:retrieval_simulated2} show that longitudinally resolved
spectral retrieval is more accurate than {\disk} 1D retrieval, as
expected. But for the {\HST}/WFC3 and {\Spitzer}/IRAC phase curves of
WASP-43b \citep{StevensonEtal2014sciWASP43bHSTphase,
StevensonEtal2017ajWASP43bSpitzerPhase} we found no significant
difference in the WASP-43b abundances between the spatially resolved
and the hemispheric integrated retrievals (\S\ref{sec:retrieval}).
This is unsurprising since the longitudinally-resolved and
hemispherically-averaged spectra overlap with each other within the
1$\sigma$ uncertainties (Figure~\ref{fig:WASP43b_stevenson_spectra}).
In other words, the uncertainties are so large for these data that
there is minimal benefit---but also no harm--- in performing a
spatially resolved retrieval over a hemispheric integration. In
Appendix~\ref{sec:analytics} we derive an analytic expression for the
longitudinal map uncertainties as a function of the uncertainties in
the spectral phase curve, and in Appendix \ref{sec:map_diffs} we
derive an expression for the difference between a phase curve and the
corresponding map, a metric for the utility of the {\resolved}
spectral retrieval approach.

\subsection{Longitudinally Resolved Spectral Retrieval of Brown Dwarfs}

While the focus on this paper has been on spatially resolved-retrieval
of transiting exoplanets, this method can in principle also be applied
to variable brown dwarfs and directly-imaged exoplanets.  Such an
analysis could reveal the properties of clouds that drive the observed
variability.  Brown dwarfs and wide-orbit companions are known to
exhibit cloud-driven variability with amplitudes as high as $\sim$25\%
in the near-infrared \citep{RadiganEtal2012apjLTbrowndwarfVariation,
BowlerEtal2020apjVHSJ1256Variability}.  However, the temperature
contrast between cloud layers in brown dwarf atmospheres is estimated
to be a few hundred
K \citep{ApaiEtal2013apjLTbrowndwarfsCloudVariability}, much smaller
than the temperature contrast between the day and night side of
short-period exoplanets. Another challenge for brown dwarfs is that
the clouds responsible for variability can evolve
rapidly \citep[e.g][]{ApaiEtal2017sciBrownDwarfWaves,
VosEtal2018mnrasABdoradusVariability} so \emph{simultaneous}
multi-band monitoring is necessary. Very few brown dwarfs were
monitored simultaneously with {\HST} and
{\Spitzer} \citep{BillerEtal2018ajPSOJ318variability,
BuenzliEtal2012apjBrownDwarfVerticalStructure}.  Our preliminary
attempts using these simultaneous data did not yield significant
differences in the longitudinally-resolved spectra.  As with the hot
Jupiter WASP-43b, we expect that observations of variable brown dwarfs
with {\JWST} may benefit from longitudinally resolved spectral
retrieval.

\section{Conclusions}
\label{sec:conclusions}

Accounting for the multi-dimensional nature of planets represents an
open challenge for the exoplanet retrieval community, and a
significant step up in complexity, relative to 1D models.  3D models
not only are more computationally demanding, but have more complex
parameter spaces that are harder to interpret.  Even model validation
becomes non trivial, while too-complex models can become easily
intractable under a retrieval framework, too-simple models might fail
to capture the inhomogeneities predicted by self-consistent physical
models.  \citet{TaylorEtal2020mnrasEmissionBiases}
and \citet{FengEtal2020aj2Dretrievals} have shown that
linear-combination models can improve the accuracy of atmospheric
retrievals under the assumption of globally constant composition and a
two-component (day and night side) temperature profile.  However,
assessing how these methods fare against a smoother longitudinal
variation of temperature and composition remains to be investigated.
In contrast, our {\resolved} approach is expected to perform better
for a planet where the physical properties vary smoothly with
longitude (as predicted by GCMs) than for a discrete two-hemispheric
model.  Furthermore, a key advantage of the {\resolved} approach is
that it provides significant gains in accuracy of the retrieved
thermal structure over {\disk} retrievals via an easy pre-processing
step to existing 1D retrieval codes.

As found here and in other multi{\hyph}dimensional studies, retrieving
accurate exoplanet compositions remains challenging.  Longitudinally
resolved retrievals provide marked improvement over {\disk}
retrievals, although both struggle with {\methane}; presumably due to
its limited impact in a spectrum dominated by {\water} features, and
its fast variation in composition between day (depleted) and night
side (plentiful).  {\citet{FengEtal2020aj2Dretrievals}}
and \citet{TaylorEtal2020mnrasEmissionBiases} found the {\methane}
abundance particularly biased when applying the traditional 1D {\disk}
analysis.  Using their inhomogeneous two-component models they
correctly recover the expected upper-limit constraint on {\methane};
however, they generated their input synthetic models assuming globally
uniform abundances (both in pressure and longitude), a simpler problem
than the ones tackled in our study and hence less liable to biases.
The longitudinal variation of minor species like {\methane} probes the
degree to which horizontal quenching can drive atmospheres towards
longitudinally uniform abundances.  Thus, properly understanding the
biases in these estimations is fundamental to assess the dynamical
properties of exoplanets.

Future missions will enable exoplanet atmospheric characterization
with unprecedented detail, but at the same time they will expose the
multiple shortcomings of present-day retrieval models that have so far
remained under the radar.  In this article we found that {\resolved}
spectral retrieval can improve the temperature and composition
estimation of inhomogeneous planets observed with the {\Webb}.

We have made available the data and analysis used in this article
at \href{https://doi.org/10.5281/zenodo.4757164}
{https://doi.org/10.5281/zenodo.4757164}

\acknowledgments

This project was conceived at the ``Multi-dimensional characterization
of distant worlds: spectral retrieval and spatial mapping'' workshop
hosted by the Michigan Institute for Research in Astrophysics and
spearheaded by Emily Rauscher.  We thank the anonymous referee for
his/her time and valuable comments.  We thank contributors to the
Python Programming Language and the free and open-source community
(see Software Section below).  We drafted this article using the
AASTeX6.2 latex template \citep{AASteamHendrickson2018aastex62}, with
further style modifications that are available
at \href{https://github.com/pcubillos/ApJtemplate}
{https://github.com/pcubillos/ApJtemplate}.  Part of this work is
based based on observations made with the NASA/ESA Hubble Space
Telescope, obtained from the data archive at the Space Telescope
Science Institute.  STScI is operated by the Association of
Universities for Research in Astronomy, Inc.\ under NASA contract NAS
5-26555.  This work is based in part on observations made with the
{\SST}, which is operated by the Jet Propulsion Laboratory, California
Institute of Technology under a contract with NASA.  This research has
made use of NASA's Astrophysics Data System Bibliographic Services.
D.\ K.\ and N.\ B.\ C.\ acknowledge support from McGill Space
Institute and the Institute for Research on Exoplanets.  E.\ G.\
acknowledges support for this work by the NSF under Grant No.\
AST-1614527 and Grant No.\ AST-1313278, by NASA under \textit{Kepler}
Grant No.\ 80NSSC19K0106, and the LSSTC Data Science Fellowship
Program, which is funded by LSSTC, NSF Cybertraining Grant No.\
1829740, the Brinson Foundation, and the Moore Foundation; her
participation in the program has benefited this work.

\software{\\
{\pyratbay} \citep{CubillosBlecic2021mnrasPyratBay},
{\mcc} \citep{CubillosEtal2017apjRednoise},
\textsc{RATE} \citep{CubillosEtal2019apjRate},
{\repack} \citep{Cubillos2017apjRepack},
\textsc{Numpy} \citep{HarrisEtal2020natNumpy},
\textsc{SciPy} \citep{VirtanenEtal2020natmeScipy},
\textsc{sympy} \citep{MeurerEtal2017pjcsSYMPY},
\textsc{Astropy} \citep{AstropyCollaboration2013aaAstropy,
 AstropyCollaboration2018ajAstropy},
\textsc{PyMC3} \citep{SalvatierEtal2016asclPyMC3},
\textsc{Matplotlib} \citep{Hunter2007ieeeMatplotlib},
\textsc{IPython} \citep{PerezGranger2007cseIPython},
AASTeX6.2 \citep{AASteamHendrickson2018aastex62},
and
\textsc{bibmanager}\footnote{
\href{http://pcubillos.github.io/bibmanager}
     {http://pcubillos.github.io/bibmanager}}
\citep{Cubillos2019zndoBibmanager}.
}

\bibliography{respect}

\appendix
\section{Analytic phase mapping}
\label{sec:analytics}

In order to most easily invert the phase variations into maps, we
would like to parameterize the wavelength-dependent phase variations
as a Fourier series, following \cite{CowanAgol2008apjPhaseInversion}:
\begin{eqnarray}
\label{good_phase}
\nonumber
F(\xi,\lambda) &=&
     F_0(\lambda) + C_1(\lambda) \cos\xi + D_1(\lambda)\sin\xi \\
 &&  + C_2(\lambda)\cos(2\xi) + D_2(\lambda)\sin(2\xi) + \ldots
\end{eqnarray}
where $\xi$ is the orbital phase from superior conjunction ($\xi=0$ at
eclipse, $\xi=\pi$ at transit). High-order harmonics are suppressed
quadratically in the phase curve (e.g., the $4\xi$ modes are 4$\times$
smaller than the $2\xi$ modes, given the same map amplitude) and the
odd harmonics ($3\xi$, $5\xi$, etc.) are invisible for an
edge-on-planet due to symmetry \citep{CowanEtal2013mnrasLightCurves},
so most authors have limited themselves to a second-order Fourier
expansion.

In order to ensure that our spatially resolved spectra have the same
normalization as the {\disk} spectra, we multiply the longitudinal map
expressions from \cite{CowanAgol2008apjPhaseInversion} by 2.  To
second order, the longitudinal planet map is therefore:
\begin{eqnarray}
\label{normalized_map}
\nonumber
\tilde{J}(\phi, \lambda) &=&
    F_0(\lambda) + \frac{4}{\pi}C_1(\lambda) \cos\phi
    - \frac{4}{\pi}D_1(\lambda)\sin\phi  \\
 && + 3 C_2(\lambda) \cos(2\phi)  -3 D_2(\lambda)\sin(2\phi),
\end{eqnarray}
where $\phi$ is the planetary longitude, with the convention that
$\phi=0$ at the sub-stellar point and increases to the East (the
direction of synchronous rotation).\footnote{These maths were derived
for an equator-on viewing geometry and hence should be accurate for a
transiting planet like WASP-43b \cite[the general solution is
presented in][]{CowanEtal2013mnrasLightCurves}.}

Omitting the wavelength dependence for clarity, the uncertainty in the
phase curve is
\begin{equation}
\label{good_phase_error}
\sigma_{F_{\xi}} = \sigma_{F_0} + \sigma_{C_1} \cos\xi + \sigma_{D_1}\sin\xi 
             + \sigma_{C_2}\cos(2\xi) + \sigma_{D_2}\sin(2\xi), 
\end{equation}
while the uncertainty in the corresponding longitudinal map is
\begin{eqnarray}
\label{good_phase_error_map}
\nonumber
\sigma_{J_{\phi}} &=&
     \sigma_{F_0} + \frac{4}{\pi}\sigma_{C_1} \cos\phi
     - \frac{4}{\pi}\sigma_{D_1}\sin\phi \\
  && + 3\sigma_{C_2}\cos(2\phi) - 3\sigma_{D_2}\sin(2\phi). 
\end{eqnarray}

\section{Difference Between the Resolved Map and the Phase Curve}
\label{sec:map_diffs}

Making the substitution $\xi = -\phi$ in Eq.~(\ref{good_phase}) yields
the disk-integrated brightness of the planet when the longitude $\phi$
is facing the observer, i.e., a phase curve that can be directly
compared to the planet's longitudinal map. The shape of the
longitudinal map differs significantly from that of the phase curve if
$|\delta(\phi, \lambda)|/\sigma_\delta (\phi)>1$, where the difference
is
\begin{eqnarray}
\label{difference}
\nonumber
\delta(\phi, \lambda) &\equiv& \tilde{J}(\phi, \lambda) - F(-\phi, \lambda) \\
\nonumber
  &=& \left(\frac{4}{\pi} -1\right)
      \Big[C_1(\lambda)\cos\phi - D_1(\lambda)\sin\phi \Big] \\
  &&  + 2\Big[ C_2(\lambda)\cos(2\phi)-D_2(\lambda)\sin(2\phi)\Big]
\end{eqnarray}
and its uncertainty is
\begin{eqnarray}
\label{difference_uncertainty}
\nonumber
\sigma_\delta^2 (\phi) &=&
     \left(\frac{4}{\pi} - 1\right)^2
     \Big[\sigma_{C_1}^2\cos^2\phi + \sigma_{D_1}^2\sin^2\phi \Big] \\
  && + 4\Big[ \sigma_{C_2}^2\cos^2(2\phi)+\sigma_{D_2}^2\sin^2(2\phi)\Big],
\end{eqnarray}
where we have again omitted the wavelength-dependence for clarity. In
other words, the longitudinal map has a significantly different shape
from the phase curve if
\begin{equation}
\label{ratio}
\frac{|\delta(\phi, \lambda)|}{\sigma_\delta (\phi)} > 1.
\end{equation}
Note the absence of the mean spectrum of the planet, $F_0(\lambda)$,
and its uncertainty, $\sigma_{F_0}$ from this expression.

In practice, spectral retrieval can only be performed on spectra, not
the difference of spectra.  In other words, uncertainties in the
eclipse spectrum, largely responsible for constraining $F_0(\lambda)$,
can make longitudinally resolved spectra consistent with {\disk}
spectra.  As such, a more salient metric is
$|\delta(\phi, \lambda)|/\sigma_\Delta (\phi)>1$, where
\begin{eqnarray}
\label{diagnostic}
\nonumber
\sigma_\Delta^2 (\phi) &=& \sigma_{F(-\phi)}^{2}+ \sigma_{\tilde{J}(\phi)}^{2} \\
\nonumber
 &=& \Big[\sigma_{F_0} + \sigma_{C_1} \cos\phi + \sigma_{D_1(\lambda)}\sin\phi \\
\nonumber
 &&  \hspace{0.2cm} +\ \sigma_{C_2}\cos(2\phi) + \sigma_{D_2}(\lambda)\sin(2\phi)\Big]^{2} \\ 
\nonumber
 &&  \Big[\sigma_{F_0} + \frac{4}{\pi}\sigma_{C_1} \cos\phi - \frac{4}{\pi}\sigma_{D_1}\sin\phi \\
 &&  \hspace{0.2cm} +\ 3\sigma_{C_2}\cos(2\phi) - 3\sigma_{D_2}(\lambda)\sin(2\phi)\Big]^{2},
\end{eqnarray}
where we have again omitted the wavelength-dependence for
clarity. Given an observed phase curve parameterized in terms of
$C_1$, $C_2$, $D_1$, $D_2$ and their uncertainties, it is a simple
matter to check whether the difference between the longitudinal map
and the lightcurve is significant.

\section{Converting Between Phase Curve Parameterizations}
\label{sec:reparameterization}

Instead of the phase curve parameterization
above, \cite{StevensonEtal2014sciWASP43bHSTphase,
StevensonEtal2017ajWASP43bSpitzerPhase} report the eclipse depth,
cosine amplitude(s), and the corresponding phase offset(s).  In
particular, for the HST/WFC3 spectrally resolved phase
curves, \cite{StevensonEtal2014sciWASP43bHSTphase} use the functional
form
\begin{equation}
F(t,\lambda) = F_0(\lambda)
    + c_1(\lambda) \cos\left(\frac{2\pi}{P}\Big[(t-t_e)-\Delta t(\lambda)\Big]\right),
\end{equation}
where $t_e$ is the time of eclipse and $\Delta t(\lambda)$ is the
phase offset from eclipse in minutes. For Spitzer/IRAC phase
curves, \cite{StevensonEtal2017ajWASP43bSpitzerPhase} adopt a second
order Fourier series and use a different convention for the phase
offset:
\begin{eqnarray}
\nonumber
F(t,\lambda) &=&
    F_0(\lambda)
       + c_1(\lambda) \cos\left(\frac{2\pi}{P}\Big[t-c_2(\lambda)\Big]\right) \\
    && + c_3(\lambda) \cos\left(\frac{4\pi}{P}\Big[t-c_4(\lambda)\Big]\right).
\end{eqnarray}

In the equations above, $t$ is time and $P$ is the planet's orbital
period. We can instead express this in terms of orbital phase, $\xi =
2\pi(t-t_e)/P$, omitting the wavelength-dependence of the coefficients
for clarity:
\begin{equation}
    F(\xi) = F_0 + c_1 \cos(\xi-\xi_1) + c_3 \cos(2(\xi-\xi_3)),
\end{equation}
where $\xi_1 = 2\pi(c_2-t_e)/P$ and $\xi_3 = 2\pi(c_4-t_e)/P$.
Using trigonometric identities, we get expressions for the Fourier
coefficients, $C_1 = c_1 \cos\xi_1$, $D_1 = c_1 \sin\xi_1$, $C_2 =
c_3 \cos(2\xi_3)$, $D_2 = c_3 \sin(2\xi_3)$, and their uncertainties:
\begin{equation}
    \sigma_{C_1}^2 = \sigma_{c_1}^2 \cos^2\xi_1 + \sigma_{\xi_1}^2 c_1^2 \sin^2\xi_1
\end{equation}
\begin{equation}
    \sigma_{D_1}^2 = \sigma_{c_1}^2 \sin^2\xi_1 + \sigma_{\xi_1}^2 c_1^2 \cos^2 \xi_1
\end{equation}
\begin{equation}
    \sigma_{C_2}^2 = \sigma_{c_3}^2 \cos^2(2\xi_3) + 4 \sigma_{\xi_3}^2 c_3^2 \sin^2(2\xi_3)
\end{equation}
\begin{equation}
    \sigma_{D_2}^2 = \sigma_{c_3}^2 \sin^2(2\xi_3) + 4\sigma_{\xi_3}^2 c_3^2 \cos^2(2\xi_3),
\end{equation}
and finally the phase offsets: $\sigma_{\xi_1}
= \frac{2\pi}{P}\sigma_{c_2}$ and $\sigma_{\xi_3}
= \frac{2\pi}{P}\sigma_{c_4}$, where we presume that the relative
uncertainty on the eclipse time and orbital period are negligible
compared to the relative uncertainty on the phase offset.

\end{document}